\documentclass[11pt,onecolumn]{IEEEtran}
\usepackage{acronym,amsmath,amssymb,amsthm,graphicx,stmaryrd,mdframed,xcolor,array}
\acrodef{AoA}{Angle of Arrival}
\acrodef{AWGN}{Additive White Gaussian Noise}
\acrodef{BER}{Bit-Error-Rate}
\acrodef{BEC}{Binary Erasure Channel}
\acrodef{BPSK}{Binary Phase-Shift Keying}
\acrodef{BSC}{Binary Symmetric Channel}
\acrodef{CDF}[CDF]{Cumulative Distribution Function}
\acrodef{CLT}[CLT]{Central Limit Theorem}
\acrodef{CSI}[CSI]{Channel State Information}
\acrodef{DMC}[DMC]{Discrete Memoryless Channel}
\acrodef{DMS}[DMS]{Discrete Memoryless Source}
\acrodef{iid}[i.i.d.]{independent and identically distributed}
\acrodef{LDPC}[LDPC]{Low-Density Parity-Check}
\acrodef{MAC}[MAC]{multiple-access channel}
\acrodef{MIMO}[MIMO]{Multiple-Input Multiple-Output}
\acrodef{MISO}{Multiple-Input Single-Output}
\acrodef{PDF}[PDF]{Probability Distribution Function}
\acrodef{PMF}[PMF]{Probability Mass Function}
\acrodef{PSD}{Power Spectral Density}
\acrodef{QPSK}{Quadrature Phase-Shift Keying}
\acrodef{SIMO}{Single-Input Multiple-Output}
\acrodef{SNR}{Signal-to-Noise Ratio}
\acrodef{wrt}[w.r.t.]{with respect to}
\acrodef{WSS}{Wide Sense Stationary}

\newcommand{\E}[2][]{{\mathbb{E}_{#1}}{\left(#2\right)}}       
\renewcommand{\P}[2][]{{\mathbb{P}_{#1}}{\left(#2\right)}}
\newcommand{\Var}[1]{{\text{\textnormal{Var}}{\left(#1\right)}}}       

\newcommand{\avgD}[2]{{{\mathbb{D}}\!\left({#1\Vert#2}\right)}}
\newcommand{\V}[1]{{{\mathbb{V}}\!\left(#1\right)}}

\newcommand{\avgI}[1]{{{\mathbb{I}}\!\left(#1\right)}}
\newcommand{\avgH}[1]{{\mathbb{H}}\!\left(#1\right)}

\newcommand{\Hb}[1]{{\mathbb{H}_b}\left(#1\right)}

\newcommand{\card}[1]{\ensuremath{\left|{#1}\right|}}           
\newcommand{\abs}[1]{\ensuremath{\left|#1\right|}}              
\newcommand{\eqdef}{\ensuremath{\triangleq}}                    
\newcommand{\intseq}[2]{\ensuremath{\llbracket{#1},{#2}\rrbracket}}  

\newcommand{\dd}{\text{\textnormal{d}}}

\renewcommand{\leq}{\leqslant}
\renewcommand{\geq}{\geqslant}

\newcommand{\proddist}{%
  \mathchoice{\raisebox{1pt}{$\displaystyle\otimes$}}
             {\raisebox{1pt}{$\otimes$}}
             {\raisebox{0.5pt}{\scalebox{0.7}{$\scriptstyle\otimes$}}}
             {\raisebox{0.4pt}{\scalebox{0.6}{$\scriptscriptstyle\otimes$}}}}
\newcommand{\pn}{{\proddist n}}

\DeclareMathAlphabet{\eurm}{U}{eur}{m}{n}
\DeclareMathAlphabet{\mathbsf}{OT1}{cmss}{bx}{n}
\DeclareMathAlphabet{\mathssf}{OT1}{cmss}{m}{sl}
\DeclareMathAlphabet{\mathcsf}{OT1}{cmss}{sbc}{n}

\DeclareSymbolFont{bsfletters}{OT1}{cmss}{bx}{n}  
\DeclareSymbolFont{ssfletters}{OT1}{cmss}{m}{n}
\DeclareMathSymbol{\bsfGamma}{0}{bsfletters}{'000}
\DeclareMathSymbol{\ssfGamma}{0}{ssfletters}{'000}
\DeclareMathSymbol{\bsfDelta}{0}{bsfletters}{'001}
\DeclareMathSymbol{\ssfDelta}{0}{ssfletters}{'001}
\DeclareMathSymbol{\bsfTheta}{0}{bsfletters}{'002}
\DeclareMathSymbol{\ssfTheta}{0}{ssfletters}{'002}
\DeclareMathSymbol{\bsfLambda}{0}{bsfletters}{'003}
\DeclareMathSymbol{\ssfLambda}{0}{ssfletters}{'003}
\DeclareMathSymbol{\bsfXi}{0}{bsfletters}{'004}
\DeclareMathSymbol{\ssfXi}{0}{ssfletters}{'004}
\DeclareMathSymbol{\bsfPi}{0}{bsfletters}{'005}
\DeclareMathSymbol{\ssfPi}{0}{ssfletters}{'005}
\DeclareMathSymbol{\bsfSigma}{0}{bsfletters}{'006}
\DeclareMathSymbol{\ssfSigma}{0}{ssfletters}{'006}
\DeclareMathSymbol{\bsfUpsilon}{0}{bsfletters}{'007}
\DeclareMathSymbol{\ssfUpsilon}{0}{ssfletters}{'007}
\DeclareMathSymbol{\bsfPhi}{0}{bsfletters}{'010}
\DeclareMathSymbol{\ssfPhi}{0}{ssfletters}{'010}
\DeclareMathSymbol{\bsfPsi}{0}{bsfletters}{'011}
\DeclareMathSymbol{\ssfPsi}{0}{ssfletters}{'011}
\DeclareMathSymbol{\bsfOmega}{0}{bsfletters}{'012}
\DeclareMathSymbol{\ssfOmega}{0}{ssfletters}{'012}

\newcommand{\calA}{{\mathcal{A}}}
\newcommand{\calB}{{\mathcal{B}}}
\newcommand{\calC}{{\mathcal{C}}}
\newcommand{\calD}{{\mathcal{D}}}

\newcommand{\calN}{{\mathcal{N}}}

\newcommand{\calP}{{\mathcal{P}}}

\newcommand{\calR}{{\mathcal{R}}}
\newcommand{\calT}{{\mathcal{T}}}
\newcommand{\calS}{{\mathcal{S}}}

\newcommand{\calX}{{\mathcal{X}}}
\newcommand{\calY}{{\mathcal{Y}}}

\newcommand{\calZ}{{\mathcal{Z}}}

\newtheorem*{remark}{Remark} 
\newtheorem{lemma}{Lemma} 
\newtheorem{corollary}{Corollary}
\newtheorem{theorem}{Theorem} 

\usepackage{mdframed}
\usepackage{microtype}

\begin{document}
\author{Matthieu R. Bloch,~\IEEEmembership{Member,~IEEE}\thanks{M. R. Bloch is with the School~of~Electrical~and~Computer~Engineering,~Georgia~Institute~of~Technology, Atlanta,~GA~30332--0250}\thanks{Parts of the results have been presented at the 2015 IEEE International Symposium on Information Theory, Hong Kong~\cite{Bloch2015}.}}
\title{Covert Communication over Noisy Channels: A Resolvability Perspective}
\date{\today}
\maketitle

\begin{abstract}
  We consider the situation in which a transmitter attempts to communicate reliably over a discrete memoryless channel while simultaneously ensuring covertness (low probability of detection) with respect to a warden, who observes the signals through another discrete memoryless channel. We develop a coding scheme based on the principle of channel resolvability, which generalizes and extends prior work in several directions. First, it shows that, irrespective of the quality of the channels, it is possible to communicate on the order of $\sqrt{n}$ reliable and covert bits over $n$ channel uses if the transmitter and the receiver share on the order of $\sqrt{n}$ key bits; this improves upon earlier results requiring on the order of $\sqrt{n}\log n$ key bits. Second, it proves that, if the receiver's channel is ``better'' than the warden's channel in a sense that we make precise, it is possible to communicate on the order of $\sqrt{n}$ reliable and covert bits over $n$ channel uses without a secret key; this generalizes earlier results established for binary symmetric channels. We also identify the fundamental limits of covert and secret communications in terms of the optimal asymptotic scaling of the message size and key size, and we extend the analysis to Gaussian channels. The main technical problem that we address is how to develop concentration inequalities for ``low-weight'' sequences; the crux of our approach is to define suitably modified typical sets that are amenable to concentration inequalities. 
\end{abstract}

\newcommand{\nll}{{\setbox0\hbox{$\ll$}\rlap{\hbox to \wd0{\hss /\hss}}\box0}}
\newcommand{\tS}{\,\smash{\widetilde{\!S}}}
\newcommand{\hS}{\,\smash{\widehat{\!S}}}
\newcommand{\hQ}{\,\vphantom{\big\|}\smash{\widehat{\!Q}}}
\newcommand{\hW}{\,\smash{\widehat{\!W}}}
\newcommand{\hT}{\,\smash{\widehat{\!T}}}
\newcommand{\hP}{\,\smash{\widehat{\!P}}}
\newcommand{\tQ}{\,\smash{\widetilde{\!Q}}}
\newcommand{\chid}[3][2]{\chi_{\raisebox{-0.2em}{\tiny $#1$}}\!\left({\left.{\!#2}\right\Vert {#3}}\right)}
\newcommand{\etad}[3][2]{\eta_{\raisebox{-0.2em}{\tiny $#1$}}\!\left({\left.{\!#2}\right\Vert #3}\right)}
\newcommand{\pL}{{\proddist L}}
\newcommand{\pell}{{\proddist \ell}}

\acrodef{KL}[KL]{Kullback-Leibler}

\definecolor{mygray}{rgb}{.95,0.9,.95}

\section{Introduction}
\label{sec:introduction}

The benefits offered by ubiquitous communication networks are now mitigated by the relative ease with which malicious users can interfere or tamper with sensitive data. The past decade has thus witnessed a growing concern for the issues of privacy, confidentiality, and integrity of communications. In many instances, users in a communication network find themselves in a position in which they wish to communicate without being detected by others. Such situations include fairly innocuous scenarios of dynamic spectrum access in wireless channels, in which secondary users attempt to communicate without being detected by primary users. A perhaps more adversarial example is a situation in which a user wishes to convey information covertly, either to maintain his privacy, avoid attacks, or escape the attention of regulatory entities monitoring the network.

Motivated by these challenges,~\cite{Bash2013,Bash2015a} have established the first characterization of the throughput at which two users may communicate reliably over a noisy channel while guaranteeing a low probability of detection from a warden, who observes the transmitted signal through another noisy channel. Specifically, it has been shown that arbitrarily low probability of detection over pure loss quantum channels, thermal noise quantum channels, and classical Gaussian channels, is possible as long as one communicates at most on the order of $\sqrt{n}$ bits over $n$ uses of the channel; this scaling result has recently been refined to establish the optimal asymptotic throughput of covert and reliable communication~\cite{Wang2015,Wang2015a}. One notable characteristic of the covert communication scheme in~\cite{Bash2013}, which we revisit in the present paper, is to require a secret key between the legitimate users with size on the order of$\sqrt{n}\log n$. These fundamental limits on covert communication may be viewed as the counterparts of the ``square root law'' of steganography~\cite{Ker2007} when the message is embedded in a covertext  with zero mean. The results of~\cite{Bash2013,Bash2015a} have been further extended in several directions, in particular by showing that arbitrarily small probability of detection is possible \emph{without secret-key} when all users are connected by \acp{BSC} and provided the warden's \ac{BSC} noise is much larger than legitimate users' \ac{BSC} noise~\cite{Che2013}; this result was also extended to include secrecy constraints~\cite{Che2014}. Other extensions have attempted to identify scenarios in which the ``square root law'' may be beaten, which includes situations in which the channel statistics are imperfectly known~\cite{Che2014a,Lee2014,Lee2014a,Lee2015}, or when the warden has uncertainty about the time of communication~\cite{bash2014,Goeckel2016}. The ideas underlying the keyless coding scheme are also connected to those developed for ``stealth'' and channel resolvability in the context of wiretap channels~\cite{Hou2014,Bloch2011e}. Tutorial presentations and discussions of these results may be found in~\cite{Che2014c,Bash2015}.

In the remainder of the paper, we use the terminology ``covert communication'' as a synonym for low-probability of detection~\cite{Hero2003,Bash2013,Bash2015a}, deniability~\cite{Che2013,Che2014a}, and undetectable communication~\cite{Lee2014,Lee2014a}, since all terms refer to the same definition. The main conceptual contribution of the present work is to revisit the problem of covert communication from the perspective of resolvability~\cite{Han1993,InformationSpectrumMethods}. This conceptual connection allows us to establish the following technical results that extend earlier work.
\begin{itemize}
\item We revisit the coding scheme of~\cite{Bash2013} that shows that on the order of $\sqrt{n}$ reliable and covert bits may be communicated over $n$ channel uses with on the order of $\sqrt{n}\log n$ bits of secret key in a universal manner; this is essentially a variation~\cite{Bash2013} with a technical refinement (Theorem~\ref{th:key-assist-covert}, Corollary~\ref{cor:scaling-key-assisted}).
\item We develop an alternative coding scheme such that, if the warden's channel statistics are known, on the order of $\sqrt{n}$ reliable covert bits may be communicated over $n$ channel uses with only on the order of $\sqrt{n}$ bits of secret key. In addition, if the legitimate user's channel is ``better'' than the warden's channel, in a sense that is made precise in Section~\ref{sec:stealth-comm-with}, we show that no secret key is needed; in particular, this generalizes~\cite{Che2013} to all \acp{DMC} (Theorem~\ref{th:key-assist-2}, Corollary~\ref{cor:scaling-conditions}).
\item We show that both the key size and the message size in our scheme are asymptotically optimal for \acp{DMC} by adapting and extending the recent converse results of Wang \textit{et al.}~\cite{Wang2015,Wang2015a} (Theorem~\ref{th:converse}, Theorem~\ref{th:general-result}, Corollary~\ref{cor:general_scaling}).
\item We extend the proposed covert communication scheme to include secrecy constraints (Theorem~\ref{th:secrecy}).
\item We partially extend the results to continuous channels, and in particular to \ac{AWGN} channels (Theorem~\ref{th:continuous_covert}).
\end{itemize}
The underlying technical problem that we solve is how to develop random coding arguments for ``low-weight'' codewords, in a sense that is precisely defined in Section~\ref{sec:techn-disgr-conc}, for which naive concentration inequalities, such as Hoeffding's inequality, do not seem to apply. The crux of our approach is to define modified ``typical sets'' that are amenable to concentration inequalities, which was inspired by an astute technique in~\cite{Che2013} to ``concentrate'' the sum of $n$ \ac{iid} random variables over a sum of $\sqrt{n}$ terms.

The paper is organized as follows. Section~\ref{sec:covert-comm-over} formally introduces the problem of covert communication, sets the notation, and establishes a few preliminary results that justify the proposed conceptual approach. Section~\ref{sec:covert-comm-with} revisits the covert communication scheme of~\cite{Bash2013} from the perspective of source resolvability, while Section~\ref{sec:stealth-comm-with} develops an alternative scheme using channel resolvability that turns out to be optimal. Section~\ref{sec:converse-results} develops the converse proof required to justify the optimality of the proposed scheme. Section~\ref{sec:extens-appl} presents several applications and extensions of the results, including Gaussian channels.

\section{Notation}
\label{sec:notation}

We briefly introduce the notation used throughout the paper. Random variables and denoted by upper case letters, e.g., $X$, while their realizations are denoted by lowercase, e.g, $x$. Vectors are denoted by boldface fonts, e.g., $\mathbf{X}$ and $\mathbf{x}$. When the length of the vector is not included as an exponent, it is implicitly assumed that vectors are of length $n\in\mathbb{N}^*$, i.e., $\mathbf{X}=(X_1,\cdots,X_n)$.

In all our calculations, $\log$ and $\exp$ are understood to the base $e$ so that the underlying unit is a nat. However, we allow ourselves to interpret and discuss our results in bits by converting $\log$ to the base two. For any $x\in\mathbb{R}$, we define $\left[x\right]^+\eqdef\max(x,0)$.

For two distributions $P$, $Q$ on some alphabet $\calX$, $\avgD{P}{Q}\eqdef\sum_x P(x)\log\frac{P(x)}{Q(x)}$ is the \ac{KL} divergence between $P$ and $Q$, and $\V{P,Q}\eqdef\frac{1}{2}\sum_x\abs{P(x)-Q(x)}$ is the total variation between $P$ and $Q$. Pinsker's inequality ensures that $\V{P,Q}^2\leq \frac{1}{2}\avgD{P}{Q}$, which we will loosen as $\V{P,Q}^2\leq \avgD{P}{Q}$ for simplicity. We say that $P$ is absolutely continuous \ac{wrt} $Q$, denoted $P\ll Q$, if for all $x\in\calX$ $P(x)=0$ if $Q(x)=0$. We also denote $P^\pn$ the product distribution $\prod_{i=1}^n P$ on $\calX^n$.

For the reader's convenience, Table~\ref{tab:notation} also provides a summary of the notation often used throughout the paper.

\begin{table}[h]
  \centering
  \caption{Commonly used notation}
  \label{tab:notation}
  \vspace{10pt}
  \begin{tabular}{c|c}
    \hline
    $\calX=\{x_0,x_1\}$& Channel input alphabet, with innocent symbol $x_0$\\
    $\{\omega_n\}_{n\geq 1}$& Indexed sequence with value in $\{0,1\}$\\
    $\alpha_n$& $\omega_n/\sqrt{n}$\\
    $P_0$& Channel output distribution $W_{Y|X=x_0}$\\
    $P_1$& Channel output distribution $W_{Y|X=x_1}$\\
    $Q_0$&  Channel output distribution $W_{Z|X=x_0}$\\
    $Q_1$&  Channel output distribution $W_{Z|X=x_1}$\\
    $\Pi_{\alpha_n}$&Channel input distribution such that $\Pi_{\alpha_n}(x_1)=\alpha_n$\\
    $P_{\alpha_n}$& Channel output distribution $P_{\alpha_n}=\alpha_n P_1+(1-\alpha_n)P_0$\\
    $Q_{\alpha_n}$&Channel output distribution $Q_{\alpha_n}=\alpha_n Q_1+(1-\alpha_n)Q_0$\\
    $\mu_0$& Minimum probability in support of $Q_0$, i.e., $\min_{z:Q_0(z)>0}Q_0(z)$\\
      \hline
  \end{tabular}
\end{table}

\section{Covert communication over noisy channels}
\label{sec:covert-comm-over}

\begin{figure}[h]
  \centering
  \includegraphics[scale=0.55]{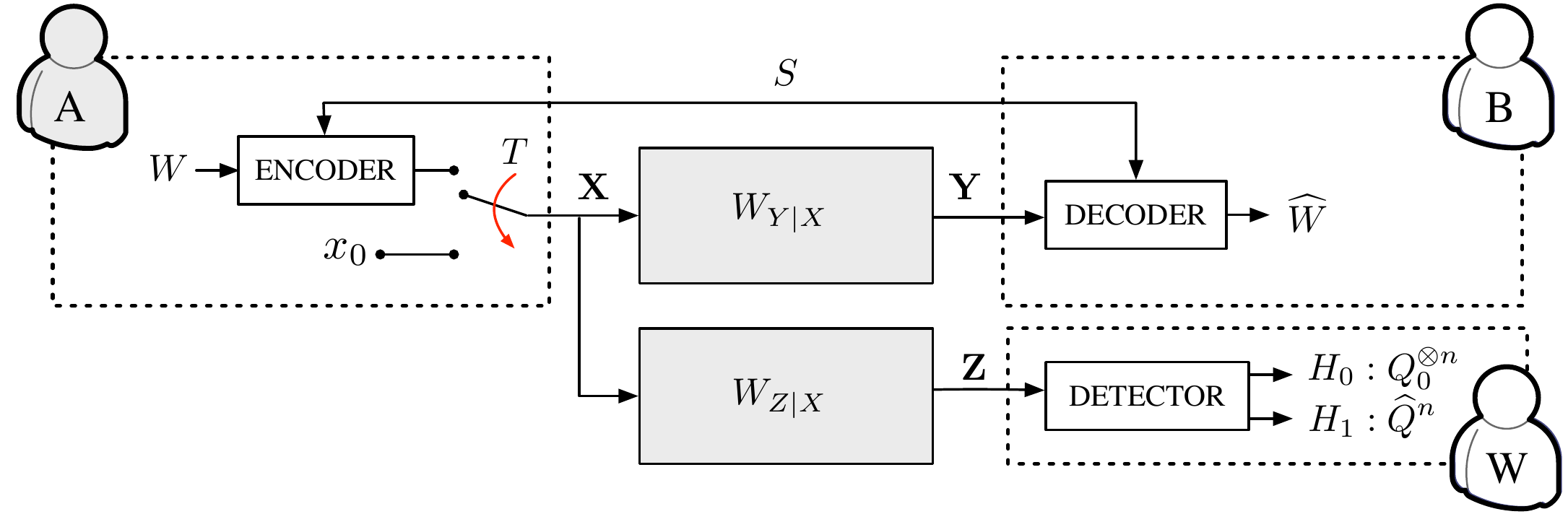}
  \caption{Model of covert communication channel.}
  \label{fig:covert_comm}
\end{figure}

We consider the situation illustrated in Fig.~\ref{fig:covert_comm}, in which two legitimate users, Alice and Bob, attempt to communicate over a \ac{DMC} $(\calX,W_{Y|X},\calY)$ without being detected by a warden, Willie, who observes the signals through another \ac{DMC} $(\calX,W_{Z|X},\calZ)$. The transition probabilities corresponding to $n$ uses of the channel are denoted $W_{Y|X}^\pn\eqdef\prod_{i=1}^nW_{Y|X}$ and $W_{Z|X}^\pn\eqdef\prod_{i=1}^nW_{Z|X}$. We also make the following assumptions.
\begin{itemize}
\item There exists an \emph{innocent symbol} $x_0\in\calX$ that corresponds to the input to the channel when no communication takes place. In such a case, the distributions induced by $x_0$ at the output of the two memoryless channels are
\begin{align}
  P_0\eqdef W_{Y|X=x_0}\quad\text{and}\quad   Q_0\eqdef W_{Z|X=x_0}\text{ with $\mu_0\eqdef \min_{z:Q_0(z)>0}Q_0(z)$}.\label{eq:p_0_q_0}
\end{align}
\item There exists another symbol $x_1\in\calX$ with $x_1\neq x_0$, and we define the distributions induced by $x_1$ at the output of the memoryless channels
\begin{align}
  P_1\eqdef W_{Y|X=x_1}\quad\text{and}\quad   Q_1\eqdef W_{Z|X=x_1}.
\end{align}
\item $Q_1\ll Q_0$ and $Q_1\neq Q_0$, which ensures that the problem is not trivial, by excluding the situations in which Willie would always detect transmission with non-vanishing probability or would never detect it. As shown in Appendix~\ref{sec:spec-cases-chann}, Alice and Bob would then communicate zero or on the order of $n$ covert bits, respectively.
\item $P_1\ll P_0$, which guarantees that Bob does not obtain an unfair advantage over Willie, by excluding the situation in which Bob could identify the location of some uncorrupted $x_1$-symbols. As shown in Appendix~\ref{sec:spec-cases-chann} Alice and Bob would then communicate on the order of $\sqrt{n}\log n$ covert bits instead of $\sqrt{n}$.
\end{itemize}
The restriction to a single symbol $x_1\neq x_0$ eases the presentation of the results, but we shall see in Section~\ref{sec:multiple-symbols} that it incurs little loss of generality. We also discuss partial extensions of the results to \ac{AWGN} channels in Section~\ref{sec:continuous-channels}. Although the absolute continuity requirements restrict the class of channels considered, they are nevertheless satisfied for large classes of channels of interest. For instance, for \ac{AWGN} channels, $x_0=0$ is the natural choice of the innocent symbol, and the absolute continuity requirements are satisfied. 

Formally, Alice's objective is to transmit a message $W$ uniformly distributed in $\intseq{1}{M}$ by encoding it into a codeword $\mathbf{X}=(X_1,\dots,X_n)$ of $n$ symbols with the help of a secret key $S$ uniformly distributed in $\intseq{1}{K}$. At the beginning of every block of $n$ symbols, Alice sets the value of a switch $T$: if $T=1$, the output of the encoder is connected to the channel; else, if $T=0$, the innocent symbol $x_0$ is sent $n$ times through the channel. Upon observing a noisy version $\mathbf{Y}=(Y_1,\dots,Y_n)$ of $\mathbf{X}$ and knowing $S$, Bob's objective is to form \emph{reliable} estimates $\hT$ and $\hW$ of $T$ and $W$, respectively. Reliability is measured by the average probability of error
\begin{align}
P_{\text{err}}\eqdef \E[S]{\P{{W\neq\hW}|S,T=1}}+\P{{\hT\neq0}|T=0}.
\end{align}
In contrast, Willie's goal is to perform a statistical test on his observation $\mathbf{Z}=(Z_1,\dots,Z_n)$ to decide whether Alice and Bob communicate (hypothesis $H_1$) or not (hypothesis $H_0$). The probability of Type I error (rejecting $H_0$ when true) is denoted $\alpha$, while the probability of Type II error (accepting $H_0$ when wrong)  is denoted $\beta$. It is possible for Willie to design blind tests that ignore his channel observations, and that achieve any pair $(\alpha,\beta)$ such that $\alpha+\beta=1$. Therefore, the objective of covert communication is to guarantee that Willie's best statistical test yields a trade-off between $\alpha$ and $\beta$ that is not much better than that of a blind test. Specifically, let $Q_0^\pn\eqdef\prod_{i=1}^nQ_0$ be the product distribution that is expected by Willie when no communication happens, and let $\hQ^n$ be the distribution expected when communication takes place. It can be shown~\cite{Lehmann2005} that Willie's optimal hypothesis test satisfies the tradeoff $\alpha+\beta\geq 1-\sqrt{\avgD{{\hQ}^n}{Q_0^\pn}}$. Therefore, achieving covert communication amounts to ensuring that $\avgD{\smash{\hQ^n}}{Q_0^\pn}$ is negligible. We provide further discussion of the role of $\avgD{\cdot}{\cdot}$ as a measure of covertness in Appendix~\ref{sec:kullb-leibl-diverg}. 

Consequently, we aim to establish scalings of $\log M$ and $\log K$ with $n$ for which there exist covert communication schemes with
\begin{align}
  \lim_{n\rightarrow\infty}P_{\text{err}}=0 \quad\text{and}\quad   \lim_{n\rightarrow\infty}\avgD{\hQ^n}{Q_0^\pn} = 0.
\end{align}

\subsection{Covert processes}
\label{sec:stealth-processes}
For $n\in\mathbb{N}^*$, let  $\alpha_n\in]0;1[$. Define the input distribution $\Pi_{\alpha_n}$ on $\{x_0,x_1\}$ such that $\Pi_{\alpha_n}(x_1)=1-\Pi_{\alpha_n}(x_0)=\alpha_n$, as well as the corresponding output distributions
\begin{align}
Q_{\alpha_n}(z)=\sum_{{x}}W_{Z|X}({z}|{x})\Pi_{\alpha_n}({x}) = Q_1(z)\alpha_n+Q_0(z)(1-\alpha_n),\label{eq:q_alpha_n}\\
P_{\alpha_n}(y)=\sum_{{x}}W_{Y|X}({y}|{x})\Pi_{\alpha_n}({x}) = P_1(y)\alpha_n+P_0(y)(1-\alpha_n).
\end{align}
Also define the product distributions 
\begin{align}
  \Pi_{\alpha_n}^\pn=\prod_{i=1}^n\Pi_{\alpha_n},\qquad Q_{\alpha_n}^\pn=\prod_{i=1}^nQ_{\alpha_n},\qquad\text{and}\qquad P_{\alpha_n}^\pn=\prod_{i=1}^nP_{\alpha_n}.\label{eq:low_weight_process}
\end{align}
Note that $Q_1\ll Q_0$ implies $Q_{\alpha_n}\ll Q_0$ and that $P_1\ll P_0$ implies $P_{\alpha_n}\ll P_0$. We then have the following result, whose proof may be found in Appendix~\ref{sec:proof-lemma}.

\begin{lemma}
  \label{lm:preliminary}
  Let $\{\alpha_n\}_{n\geq 1}$ be such that $\alpha_n\in]0;1[$ and $\lim_{n\rightarrow\infty}\alpha_n=0$. Let $Q_0$ and $Q_{\alpha_n}$ be defined as per~(\ref{eq:p_0_q_0}) and~(\ref{eq:q_alpha_n}), respectively. Define for every integer $k\geq 2$
  \begin{align}
    \chid[k]{Q_1}{Q_0}\eqdef {\sum_{z\in\calZ}\frac{\left(Q_1(z)-Q_0(z)\right)^k}{Q_0(z)^{k-1}}}\quad&\text{and}\quad \etad[k]{Q_1}{Q_0}\eqdef \sum_{z\in\calZ:Q_1(z)-Q_0(z)<0}\frac{\left(Q_1(z)-Q_0(z)\right)^k}{Q_0(z)^{k-1}}.
  \end{align}
Then, for any $n\in\mathbb{N}^*$,
\begin{align}
\avgD{Q_{\alpha_n}}{Q_0}\leq   \frac{\alpha_n^2}{2}\chid{Q_1}{Q_0} -\frac{\alpha_n^3}{6}     \chid[3]{Q_1}{Q_0} +\frac{\alpha_n^4}{3}    \chid[4]{Q_1}{Q_0}.\label{eq:upper_bound_d_covert}
\end{align}
For $n$ large enough, 
\begin{align}
  \avgD{Q_{\alpha_n}}{Q_0} \geq
\frac{\alpha_n^2}{2}\chid{Q_1}{Q_0} -\alpha_n^3\left(\frac{1}{2}\chid[3]{Q_1}{Q_0} -\frac{2}{3}\etad[3]{Q_1}{Q_0}\right)+\frac{2\alpha_n^4}{3}\etad[4]{Q_1}{Q_0}. \label{eq:lower_bound_d_covert}
\end{align}
Finally, consider the joint random variables $(X,Z)\in\{x_0,x_1\}\times\calZ$ with distribution $W_{Z|X}(z|x)\Pi_{\alpha_n}(x)$. Then,
\begin{align}
  \avgI{X;Z} = \alpha_n\avgD{Q_1}{Q_0}-\avgD{Q_{\alpha_n}}{Q_0}.\label{eq:mutual_info_lowweight}
\end{align}
\end{lemma}

\begin{remark}
The inequalities~(\ref{eq:upper_bound_d_covert}) and~(\ref{eq:lower_bound_d_covert}) may be loosened for $n$ large enough as
\begin{align}
 \frac{\alpha_n^2}{2}\chid{Q_1}{Q_0}\left(1+\sqrt{\alpha_n}\right)\geq \avgD{Q_{\alpha_n}}{Q_0} \geq  \frac{\alpha_n^2}{2}\chid{Q_1}{Q_0}\left(1-\sqrt{\alpha_n}\right).\label{eq:loose_bounds}
\end{align}
  These bounds are not tight, and one may exhibit distributions for which the inequalities are strict. Nevertheless, this allows us to obtain the correct first order and second order in $\alpha_n$ of $\avgI{X;Z}$, which is all we use in the remainder of the paper. The bounds also allow us to circumvent the rather painful Taylor series of $\avgI{X;Z}$ in $\alpha_n$.
\end{remark} 

For the specific choice $\alpha_n\eqdef\frac{\omega_n}{\sqrt{n}}$ with $\omega_n=o(1)\cap \omega(1/\sqrt{n})$ as $n\rightarrow\infty$, i.e.,\footnote{The choice of $\omega_n$ will eventually control a tradeoff between the number of covert bits and their difficulty detection by the warden. To obtain a large number of covert bits, one would choose a large $\omega_n$, say $1/\log n$. In contrast, to make the bits harder to detect, one would choose a small $\omega_n$, say $\log n/\sqrt{n}$.} 
\begin{align}
  \lim_{n\rightarrow\infty}\omega_n=0\quad\text{and}\quad   \lim_{n\rightarrow\infty}\omega_n\sqrt{n}=\infty,\label{eq:choice_omega}
\end{align}
we have
\begin{align}
  \lim_{n\rightarrow\infty}  \avgD{Q_{\alpha_n}^\pn}{Q_0^\pn}=  \lim_{n\rightarrow\infty} n\avgD{Q_{\alpha_n}}{Q_0} = 0,
\end{align}
so that $Q_{\alpha_n}^\pn$ becomes indistinguishable from $Q_0^\pn$; therefore, we call the process $Q_{\alpha_n}^\pn$ a ``covert stochastic process.'' In addition, the realizations of the input process $\Pi_{\alpha_n}$ contain an average of $\omega_n\sqrt{n}$ realizations of the $x_1$ symbol, which grows to infinity with $n$; this opens the possibility of embedding information symbols in the channel input while remaining covert. Essentially, the result of Lemma~\ref{lm:preliminary} formalizes the intuition that the change in the distribution perceived by the warden is indistinguishable from statistical noise as long as the number of $x_1$ symbols transmitted in a sequence of $n$ symbols does not exceed $\sqrt{n}$. The fact that a stochastic process with a non-trivial number of $x_1$ symbols may induce an undetectable covert stochastic process at the output of a noisy channel, suggests a generic principle for the design of covert communication schemes, which we formulate as follows. \medskip

\begin{mdframed}[backgroundcolor=mygray!40]
\centering Covert communication schemes should attempt to simulate a covert stochastic process $Q_{\alpha_n}^\pn$.
\end{mdframed}\medskip

The covert communication schemes developed in Section~\ref{sec:covert-comm-with} and Section~\ref{sec:stealth-comm-with} correspond to different applications of this principle.

\subsection{Technical digression: concentration inequalities with low-weight sequences}
\label{sec:techn-disgr-conc}
One of the technical challenges faced when trying to deal with stochastic processes such as $\Pi_{\alpha_n}^\pn$ in~\eqref{eq:low_weight_process}, is that the naive concentration inequalities traditionally used to develop information-theoretic results do not seem to apply here. To be more concrete, consider the joint random variables $(\mathbf{X},\mathbf{Z})\in\calX^n\times\calZ^n$ with the product distribution $\prod_{i=1}^n W_{Z|X}(z_i|x_i) \Pi_{\alpha_n} (x_i)$; define the mutual information random variable~\cite{InformationSpectrumMethods}
\begin{align}
\log\frac{W_{Z|X}^\pn (\mathbf{Z}|\mathbf{X})}{Q_{\alpha_n}^\pn (\mathbf{Z})}=\sum_{i=1}^n \log\frac{W_{Z|X}(Z_i|X_i)}{Q_{\alpha_n}(Z_i)},\label{eq:concentration_fails}
\end{align}
whose average is the average mutual information $\avgI{\mathbf{X};\mathbf{Z}}=n\avgI{X;Z}$. Assuming for simplicity that the range of $\log\frac{W_{Z|X}(Z_i|X_i)}{Q_{\alpha_n}(Z_i)}$ is a finite interval of length $\eta>0$,\footnote{This holds if the channel $(\calX,W_{Z|X},\calZ)$ is a fully connected DMC, such as a \ac{BSC}.} Hoeffding's inequality states that for any $\mu>0$
\begin{align}
  \P{\abs{\log\frac{W_{Z|X}^\pn (\mathbf{Z}|\mathbf{X})}{Q_{\alpha_n}^\pn (\mathbf{Z})}-n\avgI{X;Z}}\geq n\mu\avgI{X;Z}}\leq 2\exp\left(\frac{-2n\mu^2\avgI{X;Z}^2}{\eta^2}\right). \label{eq:Hoeffding_fails}
\end{align}
Unfortunately, this upper bound does not vanish because of the specific scaling of $\avgI{X;Z}$ with $n$ given in~(\ref{eq:mutual_info_lowweight}) of Lemma~\ref{lm:preliminary}. The problem finds its roots in the ``low weight'' of the sequences $\mathbf{X}$, i.e., the number of $x_1$ symbols is on average on the order of $\omega_n\sqrt{n}$, which is sub-linear in $n$. 

There are, however, some concentration inequalities that are still useful and that will be exploited in virtually all subsequent proofs. For instance, consider a \emph{binary} random sequence $\mathbf{S}\in\{0,1\}^n$ with a product distribution $\prod_{i=1}^nP_S$ such that $P_S(1)=1-P_S(0)=\frac{\omega_n}{\sqrt{n}}$. The sequence $\mathbf{S}$ is of low average weight $\omega_n\sqrt{n}$, but the application of a Chernoff bound~\cite[Exercise 2.10]{ConcentrationInequalities} yields for any $\mu\in]0;1[$
\begin{align}
  \label{eq:Chernoff_useful}
  \P{\abs{\sum_{i=1}^nS_i-\omega_n\sqrt{n}}>\mu\omega_n\sqrt{n}}\leq 2\exp\left(-\frac{\mu^2\omega_n\sqrt{n}}{3}\right),
\end{align}
which vanishes with our choice of $\omega_n$. The difference between~\eqref{eq:Chernoff_useful} and~\eqref{eq:Hoeffding_fails} may be intuitively understood as follows. The number of terms contributing to $\sum_{i=1}^nS_i$ in~\eqref{eq:Chernoff_useful} is on average $\omega_n\sqrt{n}$ because most terms are zero. In contrast, all the terms in $\sum_{i=1}^n\log\frac{W_{Z|X}(Z_i|X_i)}{Q_{\alpha_n}(Z_i)}$ are potential contributors to the sum in~\eqref{eq:Hoeffding_fails}; the concentration inequality~\eqref{eq:Hoeffding_fails} fails because the individual contributions of the terms in the sum are too small. 

We note that an alternative approach to address this technical challenge would be to use more powerful concentration inequalities, such as Bernstein's or Bennet's inequalities. We do not pursue this approach here and we rely instead on the definition of suitable typical sets, which might be of independent interest.

\section{Source-resolvability based covert communication}
\label{sec:covert-comm-with}

In this section, we revisit the architecture for covert communication proposed in~\cite{Bash2013}, which operates with a secret key $S$ of on the order of $\sqrt{n}\log n$ bits and allows the transmission of on the order of $\sqrt{n}$ bits over $n$ channel uses. The main result developed in Theorem~\ref{th:key-assist-covert} is a reinterpretation of the scheme in~\cite{Bash2013} from the perspective of source resolvability. For clarity, we assume here that $T=1$ and no attempt is made to optimize the various constants appearing in the analysis. An optimal scheme handling the general case is presented in Section~\ref{sec:stealth-comm-with}.

\begin{figure}[h]
  \centering
  \includegraphics[scale=0.55]{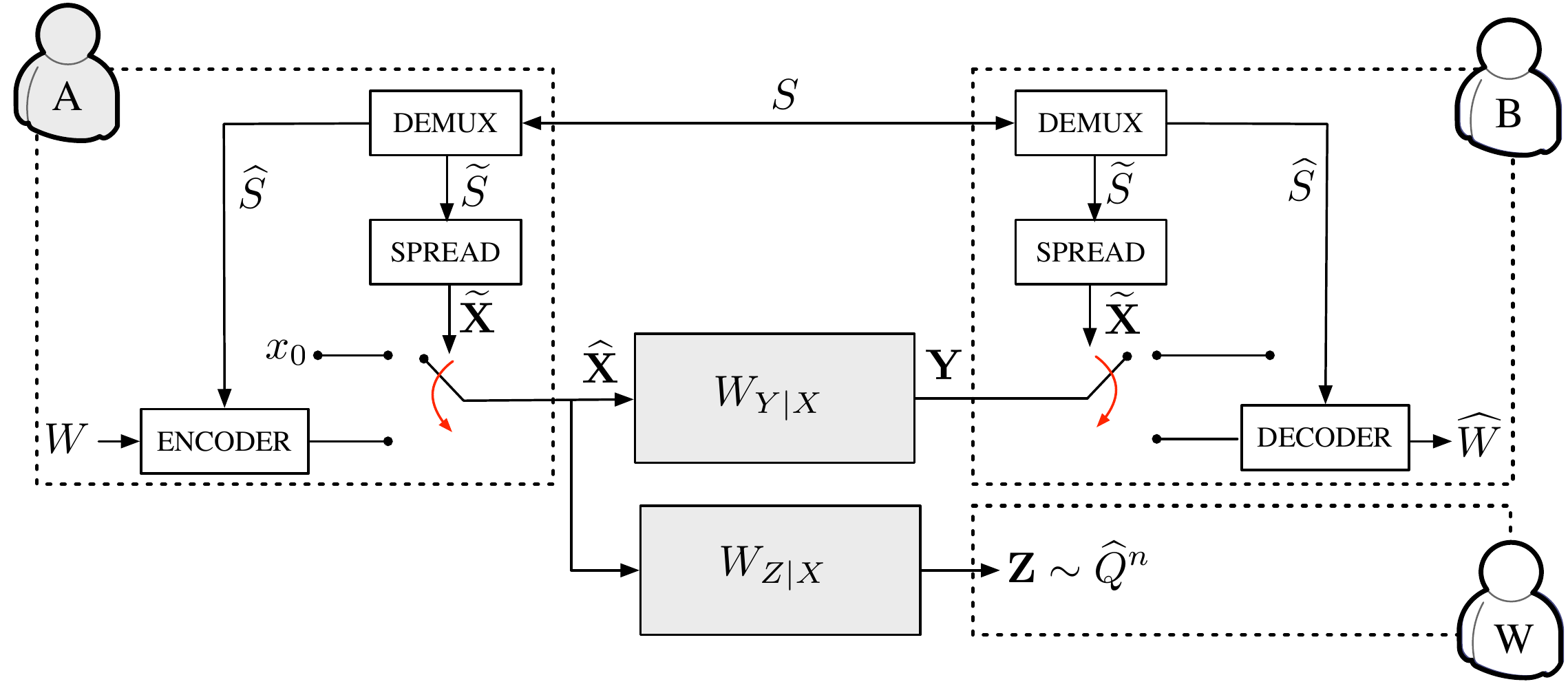}
  \caption{Covert communication scheme adapted from~\cite{Bash2013}. A secret key is used to create spreading sequences in time, which are undetectable by the warden. Information is transmitted to the legitimate receiver by modulating the spreading sequences.}
  \label{fig:universal_architecture}
\end{figure} 
The communication scheme illustrated in Fig.~\ref{fig:universal_architecture} is an adaptation to \acp{DMC} of the scheme proposed in~\cite{Bash2013} for~\ac{AWGN} channels, and it operates according to the following general principle.
\begin{enumerate}
\item Alice and Bob split the secret key $S$ into two keys $\tS\in\intseq{1}{\widetilde{K}}$ and $\hS\in\intseq{1}{\widehat{K}}$ such that $\widetilde{K}\widehat{K}=K$.
\item Alice and Bob spread the secret key $\tS$ into a length $n$ sequence $\widetilde{\mathbf{X}}\in\{x_0,x_1\}^n$.
\item Alice encodes the message $W$ into a length $n'$ binary codeword $\mathbf{B}\in\{0,1\}^{n'}$, where $n'$ on the order of $\omega_n\sqrt{n}$ will be exactly specified later.
\item Alice transmits information by modulating the symbols of $\widetilde{\mathbf{X}}$ in the position $i$ for which $\widetilde{X}_i=x_1$, resulting in a transmitted sequence $\widehat{\mathbf{X}}$. Formally, consider realizations $\widetilde{\mathbf{x}}$, $\mathbf{b}$, and $\widehat{\mathbf{s}}$, define
\begin{align}
  \text{supp}(\widetilde{\mathbf{x}})\eqdef \card{i\in\intseq{1}{n}:\widetilde{x}_i\neq x_0},
\end{align}
and let $\{i_j\}$ with $j\in\intseq{1}{  \text{supp}(\widetilde{\mathbf{x}})}$ be the positions for which $\widetilde{x}_{i_j}\neq x_0$. The symbols of the modulated sequence $\widehat{\mathbf{x}}$ are defined as
\begin{align}
  \widehat{x}_i = \left\{
  \begin{array}{l}
    x_{b_j\oplus \hat{s}_j}\text{ if $\exists j\in\intseq{1}{  \min(\text{supp}(\widetilde{\mathbf{x}}),n')}$ such that } i=i_j\\
    x_{\hat{s}_j}\text{ if $\exists j\in\intseq{\min(\text{supp}(\widetilde{\mathbf{x}}),n')}{\text{supp}(\widetilde{\mathbf{x}})}$ such that } i=i_j\\
    x_0 \text{ otherwise}.
  \end{array}\label{eq:modulation}
\right.
\end{align}
Effectively, the modulated sequence $\widehat{\mathbf{X}}$ is obtained by transmitting the sequence $\widetilde{\mathbf{X}}$ through a memoryless $Z$-channel, in which the $x_0$ symbol is unaffected and the $x_1$ symbol is flipped to the $x_0$ symbol with probability $\tfrac{1}{2}$. We denote the transition probability of this $Z$ channel by $W_{\widehat{X}|\widetilde{X}}$, and we let
\begin{align}
  W_{Z|\smash{\widetilde{X}}}(z|x) \eqdef \sum_{\hat{x}}W_{Z|X}(z|\hat{x})W_{\widehat{X}|\widetilde{X}}(\hat{x}|x).
\end{align}

\item Upon observing the channel output $\mathbf{Y}$, Bob uses his knowledge of $\widetilde{\mathbf{X}} $ to create a sequence $\widehat{\mathbf{Y}}=\left(Y_{i_1},\dots,Y_{i_{\text{supp}(\widetilde{\mathbf{X}})}}\right)$. If $\text{supp}(\widetilde{\mathbf{X}})<n'$, Bob declares an error; otherwise, it attempts to decode $\widehat{\mathbf{Y}}$ with $\widehat{S}$ to form an estimate ${\hW}$ of $W$.
\end{enumerate} 

Following the principle outlined in Section~\ref{sec:stealth-processes}, we first  attempt to simulate the process $Q_{\alpha_n}^\pn$ by simulating the {process} $\Pi_{\alpha_n}^\pn$ defined as per~(\ref{eq:low_weight_process}) at the \emph{input} of the channel.  Specifically, the secret key $\tS$ is encoded into a sequence $\widetilde{\mathbf{X}}\in\{x_0,x_1\}^n$ such that the distribution $P_{\widetilde{\mathbf{X}}}$ of $\widetilde{\mathbf{X}}$ is close to  $\Pi_{\alpha_n}^\pn$.  The following theorem characterizes the performance of this covert communication scheme. 
\begin{theorem}
  \label{th:key-assist-covert}
  Consider a discrete memoryless covert communication channel with $P_1\ll P_0$, $Q_1\ll Q_0$, and $Q_1\neq Q_0$. Let $C$ be the capacity of the main channel with inputs restricted to $\{x_0, x_1\}$ and let $\beta_n\eqdef\frac{1}{2}\frac{\omega_n}{\sqrt{n}}$ with $\omega_n\in o(1)\cap\omega(\frac{1}{\sqrt{n}})$ as $n\rightarrow\infty$. For any $\xi\in]0;1[$, there exist $\xi_1, \xi_2>0$ depending on $\xi$, $W_{Y|X}$ but not on $W_{Z|X}$, and a covert communication scheme as in Fig.~\ref{fig:universal_architecture} such that, for $n$ large enough:
  \begin{align}
    \log M=(1-\xi)\omega_n\sqrt{n}C,\quad \log K =  (1+\xi)\omega_n\sqrt{n}\log n,
  \end{align}
and
\begin{align}
  \P{W\neq {\hW}|T=1}\leq e^{-\xi_1\omega_n\sqrt{n}},\quad   \abs{  \avgD{\hQ^n}{Q_0^\pn}-\avgD{Q_{\beta_n}^\pn}{Q_0^\pn}}\leq e^{-\xi_2\omega_n\sqrt{n}}.
\end{align}
This scheme is \emph{universal} \ac{wrt} the warden's channel, in the sense that $\avgD{\hQ^n}{Q_0^\pn}$ is bounded for $n$ large enough as soon as $\chid[2]{Q_1}{Q_0}$, $\chid[3]{Q_1}{Q_0}$, and $\chid[4]{Q_1}{Q_0}$ are bounded, irrespective of the exact statistics $W_{Z|X}$.
\end{theorem}

\begin{remark}
  With some extra work, one may prove that the key $\tS$ is not necessary. Specifically, one can develop a random coding argument that includes the random generation of the code used by the legitimate users, and establish similar results without relying on a  key $\tS$. We omit the proof, which is slightly more involved but does not affect the scaling in $n$. Also note that $\tS$ acts as a one-time pad on the message, which guarantees that the message is kept confidential from the warden.
\end{remark}

\begin{IEEEproof}
  The proof of Theorem~\ref{th:key-assist-covert} consists in showing the existence of a deterministic encoder to generate $\widetilde{\mathbf{X}}$ from the key $\tS$, and the existence of a codebook with blocklength approximately $\omega_n\sqrt{n}$ to modulate $\widetilde{\mathbf{X}}$ into $\widehat{\mathbf{X}}$.
  
  \paragraph{Existence of spreading code} Let $\widetilde{K}\in\mathbb{N}^*$, $\epsilon\in]0;1[$, and define the set
  \begin{align}
    \calT_\epsilon^n&\eqdef\left\{\mathbf{x}\in\calX^n: (1-\epsilon)\omega_n\sqrt{n}\leq \text{supp}(\mathbf{x})\leq (1+\epsilon)\omega_n\sqrt{n}\right\}.\label{eq:bound_x_1_symbol}
  \end{align}
  Generate $\widetilde{K}$ codewords $\widetilde{\mathbf{x}}_i\in\{x_0,x_1\}^n$ independently at random according to the distribution
  \begin{align}
    \Pi_{\alpha_n,\epsilon}^n(\mathbf{x}) \eqdef \frac{\mathbf{1}\{\mathbf{x}\in\calT_\epsilon^n\}}{\lambda_n}\Pi_{\alpha_n}^\pn(\mathbf{x}) \quad\text{with }\alpha_n\eqdef\frac{\omega_n}{\sqrt{n}}\text{ and }\lambda_n\eqdef \P[\Pi_{\alpha_n}^\pn]{\mathbf{X}\in\calT_\epsilon^n}.\label{eq:truncated_dist}
  \end{align}
  Using a Chernoff bound, we have
  \begin{align}
    1-\lambda_n=\P[\Pi_{\alpha_n}^\pn]{\mathbf{X}\notin\calT_\epsilon^n}\leq 2e^{-\frac{1}{3}\epsilon^2\omega_n\sqrt{n}}\label{eq:bound_lambda}.
  \end{align}
  Finally, define the output distribution corresponding to $\Pi_{\alpha_n,\epsilon}^n$ as
  \begin{align}
    Q_{\alpha_n,\epsilon}^n\eqdef\sum_{\mathbf{x}}W_{Z|X}^\pn(\mathbf{z}|\mathbf{x}) \Pi_{\alpha_n,\epsilon}^n(\mathbf{x}).\label{eq:def_output_truncdist}
  \end{align}

  The encoder spreads a secret key $\tilde{s}\in\intseq{1}{\widetilde{K}}$ into a sequence $\widetilde{\mathbf{x}}\in\{x_0,x_1\}^n$ according to the map $\intseq{1}{\widetilde{K}}\rightarrow\{x_0,x_1\}^n:\tilde{s}\mapsto \widetilde{\mathbf{x}}_{\tilde{s}}$. The resulting spreading sequence distribution is then
  \begin{align}
    P_{\widetilde{\mathbf{X}}}(\mathbf{x})=\sum_{i=1}^{\widetilde{K}}\frac{1}{\widetilde{K}}\mathbf{1}\left\{\mathbf{x}=\widetilde{\mathbf{x}}_i\right\}.
  \end{align}
Our objective is to show that for suitably large $\widetilde{K}$, the spreading sequence distribution $P_{\widetilde{\mathbf{X}}}$ is close to the product distribution $\Pi_{\alpha_n}^\pn$. This is actually a variation of \emph{source resolvability}~\cite{Han1993}, which we detail to carefully handle the dependence of $\Pi_{\alpha_n}^\pn$ on $\alpha_n$. As shown in Appendix~\ref{sec:proof-source-resolvability-div}, the average of $\avgD{P_{\widetilde{\mathbf{X}}}}{\Pi_{\alpha_n}^\pn}$ over the random code generation satisfies the following.
\begin{lemma}
  \label{lm:source_resolvability_div}
  For any $\gamma>0$ and all $n\in\mathbb{N}^*$ large enough,
\begin{align}
  \E{\avgD{P_{\widetilde{\mathbf{X}}}}{\Pi_{\alpha_n}^\pn} }\leq \frac{n}{\lambda_n}\log\left(\frac{2}{\alpha_n}\right)\P[\Pi_{\alpha_n}^\pn]{\textnormal{supp}(\mathbf{X})\geq\frac{\gamma+n\log (1-\alpha_n)}{\log \frac{1-\alpha_n}{\alpha_n}}} + \log\left(\frac{1}{\lambda_n}+\frac{e^\gamma}{\widetilde{K}}\right).
\end{align}
\end{lemma}
For any $\mu>0$, by choosing
  \begin{align}
    \gamma = (1+\mu)\omega_n\sqrt{n}\log \left(\frac{1}{\alpha_n}-1\right) -n{\log (1-\alpha_n)}
  \end{align}
  and noticing that $\text{supp}(\mathbf{X})=\sum_{i=1}^n\mathbf{1}\{X_i=x_1\}$ with $\E[\Pi_{\alpha_n}^\pn]{\text{supp}(\mathbf{X})}=\omega_n\sqrt{n}$, we obtain with a Chernoff bound
  \begin{align}
    \P[\Pi_{\alpha_n}^\pn]{\text{supp}(\mathbf{X})\geq\frac{\gamma+n\log (1-\alpha_n)}{\log \frac{1-\alpha_n}{\alpha_n}}} = \P[\Pi_{\alpha_n}^\pn]{\text{supp}(\mathbf{X})>(1+\mu)\omega_n\sqrt{n}}\leq e^{-\frac{1}{3}\mu^2\omega_n\sqrt{n}}.
  \end{align}
  With $\alpha_n=\frac{\omega_n}{\sqrt{n}}$ as per~(\ref{eq:choice_omega}), notice that
  \begin{align}
    \gamma &=(1+\mu)\omega_n\sqrt{n}\log \left(\frac{1}{\alpha_n}-1\right) -n{\log (1-\alpha_n)}\leq (1+\mu)\omega_n\sqrt{n}\left(\log  \sqrt{n}-\log {\omega_n}\right) +\frac{n\omega_n}{\sqrt{n}-\omega_n}
  \end{align}
  where we have used the inequality $\log  (1+x)\geq \frac{x}{1+x}$ for $x\in]-1;\infty[$. For $n$ large enough, we also have $\log  \sqrt{n}-\log \omega_n<\log {n}$ by~(\ref{eq:choice_omega}) and $\sqrt{n}-\omega_n\geq\frac{\sqrt{n}}{\mu\log  n}$, so that $ \gamma \leq (1+2\mu)\omega_n\sqrt{n}\log  n$ and $\log\frac{2}{\alpha_n}\leq \log 2+\log n$. Hence, choosing
  \begin{align}
    \log  \widetilde{K} = (1+\delta) (1+2\mu)\omega_n\sqrt{n}\log  n\quad\text{ with any }\delta>0,\label{eq:key_scheme_size_k_tilde}
  \end{align}
  we obtain for $n$ large enough
  \begin{align}
    \E{\avgD{P_{\widetilde{\mathbf{X}}}}{\Pi_{\alpha_n}^\pn}}      &\leq e^{-\rho \omega_n\sqrt{n}} \text{ with some appropriately defined } \rho>0.\label{eq:source_resolvability}
  \end{align}
 In particular, there exists a specific code for which
 \begin{align}
 \avgD{P_{\widetilde{\mathbf{X}}}}{\Pi_{\alpha_n}^\pn} \leq e^{-\rho \omega_n\sqrt{n}}\quad\text{and}\quad \V{P_{\widetilde{\mathbf{X}}},\Pi_{\alpha_n}^\pn}\leq e^{-\frac{1}{2}\rho \omega_n\sqrt{n}}  \label{eq:bound_div_source_res},
 \end{align}
where the bound on $\V{\cdot,\cdot}$ follows by Pinsker's inequality.

\paragraph{Effect of modulation} Irrespective of the error-control code used to encode $W$, modulation requires at most
\begin{align}
  \log  \widehat{K} = (1+\epsilon)\omega_n\sqrt{n}\label{eq:key_scheme_size_k_hat}
\end{align}
by the constraint imposed in~(\ref{eq:bound_x_1_symbol}), which is negligible compared to $\log\widetilde{K}$ in~(\ref{eq:key_scheme_size_k_tilde}). When presenting the distribution $\Pi_{\alpha_n}$ at the input of the $Z$-channel $W_{\widehat{X}|\widetilde{X}}$ induced by the modulation, one may check that the corresponding distribution at the output of the $Z$-channel is $\Pi_{{\beta_n}}$ with $\beta_n\eqdef \frac{\alpha_n}{2}$. Consequently, we have by the data processing inequality
\begin{align}
  \avgD{\hQ^n}{Q_{\beta_n}^\pn} \leq \avgD{P_{\widetilde{\mathbf{X}}}}{\Pi_{\alpha_n}^\pn}\quad\text{and}\quad\V{\hQ^n,Q_{\beta_n}^\pn}\leq \V{P_{\widetilde{\mathbf{X}}},\Pi_{\alpha_n}^\pn}.
\end{align}
Next, notice that
\begin{align}
  \avgD{\hQ^n}{Q_0^\pn}  &=  \avgD{\hQ^n}{Q_{\beta_n}^\pn} + \sum_{\mathbf{z}}\hQ^n(\mathbf{z})\log\frac{Q_{\beta_n}^\pn (\mathbf{z})}{Q_0^{\pn}(\mathbf{z})}\\
  &= \avgD{\hQ^n}{Q_{\beta_n}^\pn} +\avgD{Q_{\beta_n}^\pn}{Q_0^\pn} + \sum_{\mathbf{z}}\left(\hQ^n(\mathbf{z})-Q_{\beta_n}^\pn(\mathbf{z})\right)\log\frac{Q_{\beta_n}^\pn (\mathbf{z})}{Q_0^{\pn}(\mathbf{z})},\label{eq:bound_div_source_res_2}
\end{align}
with
\begin{align}
  \abs{ \sum_{\mathbf{z}}\left(\hQ^n(\mathbf{z})-Q_{\beta_n}^\pn(\mathbf{z})\right)\log\frac{Q_{\beta_n}^\pn (\mathbf{z})}{Q_0^{\pn}(\mathbf{z})}}\leq n\V{\hQ^n,Q_{\beta_n}^\pn}\log\frac{1}{\mu_0}\leq n e^{-\frac{1}{2}\rho \omega_n\sqrt{n}}\log\frac{1}{\mu_0}.\label{eq:bound_div_source_res_3}
\end{align}
Hence, combining~(\ref{eq:bound_div_source_res})-(\ref{eq:bound_div_source_res_3}), we conclude that there exists a constant $\xi_2>0$ such that, for $n$ large enough,
\begin{align}
  \abs{  \avgD{\hQ^n}{Q_0^\pn}-\avgD{Q_{\beta_n}^\pn}{Q_0^\pn}}\leq e^{-\xi_2\omega_n\sqrt{n}}.\label{eq:bound_div_source_res_4}
\end{align}

\paragraph{Reliability} We conclude the proof by showing how one may encode the messages $W$ into codewords $\mathbf{B}$. Assume that the main channel has capacity $C$ when inputs are restricted to the set $\{x_0,x_1\}$. Standard arguments~\cite{TopicsMultiUserIT} show that, for any $\delta>0$, there exists a binary code of length $(1-\epsilon)\omega_n\sqrt{n}$, such that one may choose
\begin{align}
 \log  M = (1-\delta)(1-\epsilon)\omega_n\sqrt{n} C \label{eq:key_scheme_size_m}
\end{align}
with probability of error
\begin{align}
  P_{\text{err}}\leq e^{-\xi_1\omega_n\sqrt{n}},\label{eq:key_scheme_bound_perr}
\end{align}
where $\xi_1>0$ depend on $\delta$, $\epsilon$, and $W_{Y|X}$. 

Combining the choice of $\log  \widehat{K}$, $\log  \widetilde{K}$, and $\log  M$, in~(\ref{eq:key_scheme_size_k_tilde}),~(\ref{eq:key_scheme_size_k_hat}),~(\ref{eq:key_scheme_size_m}), with the bounds obtained in~(\ref{eq:bound_div_source_res_4}) and~(\ref{eq:key_scheme_bound_perr}), one may then find the appropriate constant $\xi$ promised in the statement of the theorem.
\end{IEEEproof}

The interpretation of Theorem~\ref{th:key-assist-covert} is the same as in~\cite[Theorem 1.2]{Bash2013}. However, our underlying covert communication scheme is slightly different, as the key $S$ may be viewed as the seed to generate a ``spreading sequence,'' rather than a way to index the positions for transmission. Technically, the result also differs from~\cite{Bash2013} by ensuring a bound on the maximum key size instead on the average key size, although it is still on the order of $\sqrt{n} \log n$. Finally, Theorem~\ref{th:key-assist-covert} relies on more sophisticated resolvability techniques~\cite{InformationSpectrumMethods}, whose usefulness will become apparent in Section~\ref{sec:stealth-comm-with}. 

From Theorem~\ref{th:key-assist-covert}, one may now attempt to establish asymptotic limits akin to capacity. Unlike traditional information theoretic problems, there seems to be no strong converse and the factor $\omega_n$ that controls the decay of $\avgD{\hQ^n}{Q_0^\pn}$ also affects $\log M$. Consequently, following the approach of~\cite{Wang2015a}, $\log M$ is scaled by $\sqrt{n\avgD{{\hQ}^n}{Q_0^\pn}}$ to obtain a meaningful asymptotic constant. 
\begin{corollary}
\label{cor:scaling-key-assisted}
  Consider a discrete memoryless covert communication channel with $P_1\ll P_0$, $Q_1\ll Q_0$, and $Q_1\neq Q_0$. Let $C$ be the capacity of the main channel with inputs restricted to $\{x_0, x_1\}$. For any $\xi\in]0;1[$, there exist covert communication schemes such that
  \begin{align*}
    \lim_{n\rightarrow\infty}\avgD{\hQ^n}{Q_0^\pn}=0, &\quad \lim_{n\rightarrow\infty}{\P{W\neq{\hW}|T=1}}=0,
  \end{align*}
  and
\begin{align*}
  \lim_{n\rightarrow\infty}\frac{\log M}{\sqrt{n\avgD{{\hQ}^n}{Q_0^\pn}}} =  2(1-\xi)\sqrt{\frac{2}{\chid[2]{Q_1}{Q_0}}}C,&\quad
                                                                                                                                   \lim_{n\rightarrow\infty}\frac{\log K}{\sqrt{n\avgD{{\hQ}^n}{Q_0^\pn}}} =  \infty.
  \end{align*}
\end{corollary}
\begin{IEEEproof}
    Consider a sequence of coding schemes as identified by Theorem~\ref{th:key-assist-covert} for some $\xi>0$. Then, using the remark after Lemma~\ref{lm:preliminary}, we have
  \begin{align}
    \avgD{\hQ^n}{Q_{0}^\pn}&\leq n\avgD{Q_{\beta_n}}{Q_{0}} + e^{-\xi_2\omega_n\sqrt{n}}\leq \frac{\omega_n^2}{8}\chid{Q_1}{Q_0}\left(1+\sqrt{\frac{\omega_n}{2\sqrt{n}}}\right) + e^{-\xi_2\omega_n\sqrt{n}},\\
    \avgD{\hQ^n}{Q_{0}^\pn}&\geq n\avgD{Q_{\beta_n}}{Q_{0}} - e^{-\xi_2\omega_n\sqrt{n}}\geq \frac{\omega_n^2}{8}\chid{Q_1}{Q_0}\left(1-\sqrt{\frac{\omega_n}{2\sqrt{n}}}\right) - e^{-\xi_2\omega_n\sqrt{n}}.
  \end{align}
Hence, $\lim_{n\rightarrow\infty}\avgD{\hQ^n}{Q_0^\pn}=0$, and using the constraints on $\omega_n$ as per~(\ref{eq:choice_omega}), we obtain
\begin{align}
      \lim_{n\rightarrow\infty} \frac{\log M}{\sqrt{n\avgD{{\hQ}^n}{Q_0^\pn}}} &\geq       \lim_{n\rightarrow\infty} \frac{(1-\xi)\omega_n\sqrt{n}C}{\omega_n\sqrt{n}\sqrt{ \frac{1}{8}\chid{Q_1}{Q_0}(1+\frac{\omega_n}{2\sqrt{n}}) + \frac{1}{\omega_n^2}e^{-\xi_2\omega_n\sqrt{n}}}} \nonumber\\
                                                                               &= 2\sqrt{\frac{2}{\chid{Q_1}{Q_0}}}(1-\xi)C,
\end{align}
\begin{align}
  \lim_{n\rightarrow\infty} \frac{\log M}{\sqrt{n\avgD{{\hQ}^n}{Q_0^\pn}}} &\leq       \lim_{n\rightarrow\infty} \frac{(1-\xi)\omega_n\sqrt{n}C}{\omega_n\sqrt{n}\sqrt{ \frac{1}{8}\chid{Q_1}{Q_0}(1-\frac{\omega_n}{2\sqrt{n}}) - \frac{1}{\omega_n^2}e^{-\xi_2\omega_n\sqrt{n}}}}\nonumber\\
  &= 2\sqrt{\frac{2}{\chid{Q_1}{Q_0}}}(1-\xi)C.
\end{align}
\end{IEEEproof}

\section{Channel-resolvability based covert communication}
\label{sec:stealth-comm-with}

\begin{figure}[h]
  \centering
  \includegraphics[scale=0.55]{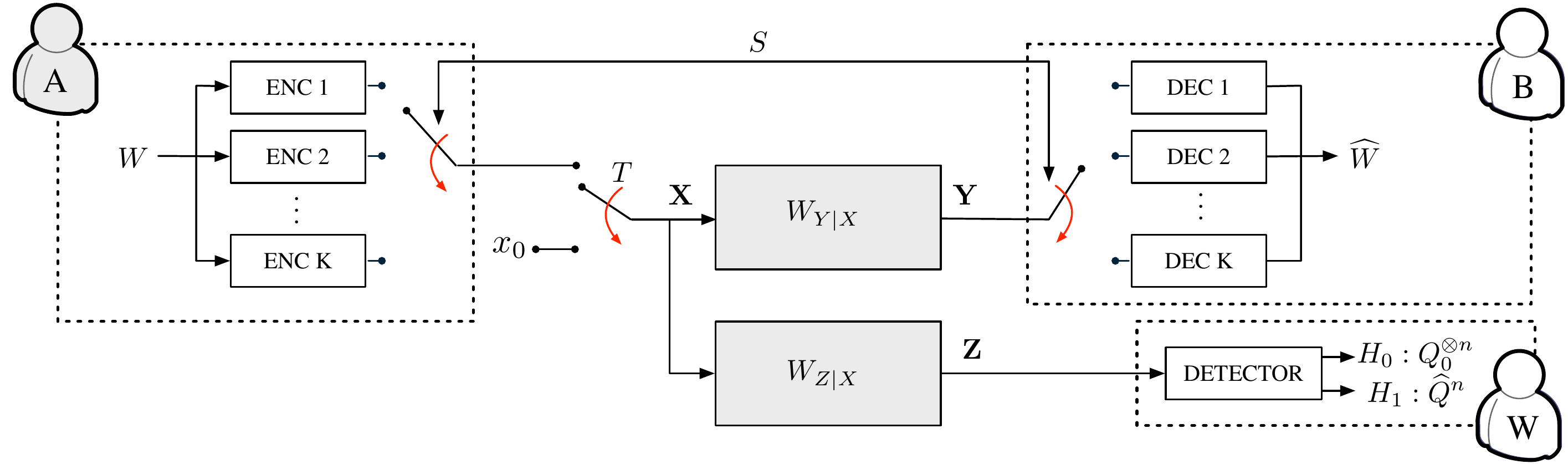}
  \caption{Channel-resolvability based covert communication. The key $S$ is used to select one of $K$ possible codebooks.}
  \label{fig:channel-resol-covert}
\end{figure}

The covert communication scheme analyzed in Theorem~\ref{th:key-assist-covert} requires a key size on the order of $\omega_n\sqrt{n}\log n$ bits to transmit on the order of $\omega _n \sqrt{n}$ bits. In a practical implementation of the coding scheme, the key is likely to stem from a pseudo-random number generator, which opens the proposed scheme to attacks that could get particularly detrimental as the required key gets longer. Fortunately, we show next how the scheme may be suitably modified to use a key size on the order of $\omega_n\sqrt{n}$ bits. The idea behind the improvement is to use the key $S$ to help \emph{directly} simulate the covert process $Q_{\alpha_n}^\pn$ defined as per~(\ref{eq:low_weight_process}), without simulating the process $\Pi_{\alpha_n}^\pn$. Conceptually, the idea is to rely on \emph{channel resolvability} in place of \emph{source resolvability}, but the precise analysis requires some care because of the ``low weight'' nature of the process $\Pi_{\alpha_n}^\pn$. The use of channel resolvability also enables one to improve upon the value of $\log M$ in Theorem~\ref{th:key-assist-covert}. The proposed architecture is illustrated in Fig.~\ref{fig:channel-resol-covert}. The key $S$ is used to select one of $K$ codebooks, each containing $M$ codewords for encoding message $W$. The underlying idea is then to guarantee that each codebook is sufficiently small to ensure reliability over the main channel while ensuring that there are sufficiently many distinct codewords overall to keep the warden confused.

\begin{theorem}
  \label{th:key-assist-2}
    Consider a discrete memoryless covert communication channel with $P_1\ll P_0$, $Q_1\ll Q_0$, and $Q_1\neq Q_0$. Let $\alpha_n\eqdef\frac{\omega_n}{\sqrt{n}}$ with $\omega_n\in o(1)\cap\omega(\frac{1}{\sqrt{n}})$ as $n\rightarrow\infty$. For any $\xi\in]0;1[$, there exist $\xi_1, \xi_2>0$ depending on $\xi$, $W_{Y|X}$, $W_{Z|X}$, and a covert communication scheme as in Fig.~\ref{fig:channel-resol-covert} such that, for $n$ large enough,
  \begin{align*}
    \log M &= (1-\xi)\omega_n\sqrt{n} \avgD{P_1}{P_0},\\
    \log K &= \omega_n\sqrt{n}\left[ (1+\xi) \avgD{Q_1}{Q_0}-(1-\xi)\avgD{P_1}{P_0}\right]^+,
  \end{align*}
  and
\begin{align*}
  P_{\text{err}}\leq e^{-\xi_1\omega_n\sqrt{n}},\qquad    \abs{\avgD{\hQ^n}{Q_{0}^\pn}-\avgD{Q_{\alpha_n}^\pn}{Q_{0}^\pn}} \leq e^{-\xi_2\omega_n\sqrt{n}}.
  \end{align*}
\end{theorem}
\begin{remark}
  The proof of Theorem~\ref{th:key-assist-2} actually shows an exponential concentration result, in the sense that a randomly generated codebook satisfies the reliabilty and covertness conditions with  probability at least $1-e^{-\theta\omega_n\sqrt{n}}$ for some $\theta>0$. In some cases, it is possible to strengthen the result and show a \emph{super-exponential } concentration result~\cite{Che2014}, in the sense that a randomly generated codebook satisfies  the reliability and covertness conditions with  probability at least $1-e^{-e^{\theta \omega_n\sqrt{n}}}$ for some $\theta>0$.
\end{remark}

Notice that no key is needed if $\avgD{P_1}{P_0}>\avgD{Q_1}{Q_0}$ by choosing $\xi$ small enough, in which case a single codebook ($K=1$) is sufficient to achieve both resolvability and reliability simultaneously. In contrast, when $\avgD{P_1}{P_0}\leq \avgD{Q_1}{Q_0}$, the proposed scheme requires a key to achieve covert communication. 

\begin{IEEEproof}
  The proof of Theorem~\ref{th:key-assist-2} is essentially a random coding argument for channel reliability~\cite{Verdu1994} and channel resolvability~\cite{Cuff2013}; however, because the number of bits communicated is on the order of $\omega_n\sqrt{n}$ over $n$ channel uses, naive concentration inequalities do not seem to apply directly. The idea we exploit to circumvent this technical issue is to use suitably modified typical sets.

\paragraph{Random codebook generation} Let $M,K\in\mathbb{N}^*$. Generate $MK$ codewords $\mathbf{x_{ij}}\in\{x_0,x_1\}^n$ with $i\in\intseq{1}{M}$ and $j\in\intseq{1}{K}$ independently according to the product distribution $\Pi_{\alpha_n}^\pn$. Define the set\footnote{The traditional typical set for decoding is similar to $\calA^n_\gamma$ but with $P_{\alpha_n}^\pn(\mathbf{y)}$ in place of $P_0^\pn(\mathbf{y)}$~\cite{Verdu1994}. This amounts to using the information density in place of the relative entropy density.}
  \begin{align}
    \calA^n_\gamma&\eqdef\left\{(\mathbf{x},\mathbf{y})\in\calX^n\times\calY^n:\log \frac{W_{Y|X}^\pn(\mathbf{y}|\mathbf{x})}{P_0^\pn(\mathbf{y})}>\gamma\right\}
  \end{align}
  where $\gamma>0$ will be determined later. The encoder simply maps a message $i$ and a key $j$ to the codeword $\mathbf{x}_{ij}$. The decoder, which has access to $\mathbf{y}$ and the key $j$, operates as follows:
  \begin{itemize}
  \item if there exists a unique $i\in\intseq{1}{M}$ such that $(\mathbf{x}_{ij},\mathbf{y})\in\calA^n_\gamma$, output $\hT=1$ and $\hW=i$;
  \item if there is no codeword $i$ such that $(\mathbf{x}_{ij},\mathbf{y})\in\calA^n_\gamma$, declare there was no communication and $\hT=0$;\footnote{Since the scheme sometimes allows keyless operation ($K=1$), the decoder must be able to identify the absence of transmission without relying on a key.}
  \item otherwise, declare a decoding error.
  \end{itemize}

  \paragraph{Channel reliability analysis} As shown in Appendix~\ref{sec:proof-lemma-reliability-lpd}, the probability of decoding error $P_{\text{err}}$ averaged of the random codebook satisfies the following.
  \begin{lemma}
    \label{lm:reliability-lpd}
    For any $\gamma>0$,
    \begin{align}
      \E{P_{\text{err}}}\leq \P[W_{Y|X}^\pn\Pi_{\alpha_n}^\pn]{\log \frac{W_{Y|X}^\pn(\mathbf{Y}|\mathbf{X})}{P_0^\pn(\mathbf{Y})}\leq\gamma} + M e^{-\gamma}(1+\exp\left(\omega_n^2(\zeta-1)\right))\quad\text{with}\quad\zeta\eqdef\sum_{y}\frac{P_1(y)^2}{P_0(y)}.\label{eq:bound_pe_one_shot}
    \end{align}
  \end{lemma}
  We now analyze the first term on the right-hand side of~(\ref{eq:bound_pe_one_shot}) more precisely. Since $W_{Y|X}^\pn$ and $\Pi_{\alpha_n}^\pn$ are product distributions,
  \begin{align}
\P[W_{Y|X}^\pn\Pi_{\alpha_n}^\pn]{\log \frac{W_{Y|X}^\pn(\mathbf{Y}|\mathbf{X})}{P_0^\pn(\mathbf{y})}\leq\gamma} &= \P[W_{Y|X}^\pn\Pi_{\alpha_n}^\pn]{\sum_{i=1}^n\log \frac{W_{Y|X}(Y_i|X_i)}{P_0(Y_i)}\leq\gamma}.\label{eq:bound_iid_sum}
  \end{align}
  If $X_i=x_0$, note that $Y_i$ is distributed according to $P_0$ and that $\log \frac{W_{Y|X}(Y_i|x_0)}{P_0(Y_i)}=0$. Similarly, if $X_i=x_1$, $Y_i$ is distributed according to $P_1$ and $\log \frac{W_{Y|X}(Y_i|x_1)}{P_0(Y_i)}=\log \frac{P_1(Y_i)}{P_0(Y_i)}$. Consequently, although the sum in~(\ref{eq:bound_iid_sum}) contains $n$ terms, only those for which $X_i=x_1$ contribute to it. Therefore, we introduce the random variable  $L\eqdef \sum_{i=1}^n\mathbf{1}\left\{X_i=x_1\right\}$, so that
  \begin{align}
    \P[W_{Y|X}^\pn\Pi_{\alpha_n}^\pn]{\sum_{i=1}^n\log \frac{W_{Y|X}(Y_i|X_i)}{P_0(Y_i)}\leq\gamma} &=   \E[L]{\left.\P[W_{Y|X}^\pn\Pi_{\alpha_n}^\pn]{\sum_{i=1}^n\log \frac{W_{Y|X}(Y_i|X_i)}{P_0(Y_i)}\leq\gamma}\right\vert L}\nonumber\\
                                                                                           &=   \E[L]{\left.\P[P_1^{\pL}]{\sum_{i=1}^{L}\log \frac{P_1(Y_i)}{P_0(Y_i)}\leq\gamma}\right\vert L}\label{eq:reliability_concentration_1}
  \end{align}
  Let $\mu,\nu\in]0;1[$ and set
  \begin{align}
    \gamma\eqdef (1-\mu)(1-\nu)\omega_n\sqrt{n}\avgD{P_1}{P_0}\quad\text{and}\quad\calC_\mu^n\eqdef\{\ell\in\mathbb{N}^*:\ell>(1-\mu)\omega_n\sqrt{n}\}.
  \end{align}
  Intuitively, $\exp\gamma$ represents the number of codewords in a codebook while $\calC_\mu^n$ represents the likely support size of the codewords. Then,
  \begin{align}
    \E[L]{\left.\P[P_1^\pL]{\sum_{i=1}^{L}\log \frac{P_1(Y_i)}{P_0(Y_i)}\leq\gamma}\right\vert L}\leq \sum_{\ell\in \calC^n_\mu}\P{L=\ell}\P[P_1^\pell]{\sum_{i=1}^{\ell}\log \frac{P_1(Y_i)}{P_0(Y_i)}\leq\gamma} + \P{L\notin \calC_\mu^n}
    \label{eq:reliability_concentration_2}
  \end{align}
  Since $\E{L} = \sum_{i=1}^n\E{\mathbf{1}\left\{X_i=x_1\right\}}=\omega_n\sqrt{n}$, we obtain with a Chernoff bound
  \begin{align}
    \P{L\notin \calC_\mu^n} = \P{L\leq (1-\mu)\E{L}} \leq e^{-\frac{1}{2}\mu^2\omega_n\sqrt{n}}. \label{eq:reliability_concentration_3}
  \end{align}
  For $\ell\in\calC_\mu^n$, we have
  \begin{align}
    (1-\mu)(1-\nu)\omega_n\sqrt{n} -\ell <   (1-\nu)\ell - \ell = -\nu \ell
  \end{align}
  so that
  \begin{align}
    \P[P_1^\pell]{\sum_{i=1}^{\ell}\log \frac{P_1(Y_i)}{P_0(Y_i)}\leq\gamma} &=\P[P_1^\pell]{\sum_{i=1}^{\ell}\log \frac{P_1(Y_i)}{P_0(Y_i)}-\ell\avgD{P_1}{P_0}\leq\gamma-\ell\avgD{P_1}{P_0}}\nonumber\\
                                                               &\leq \P[P_1^\pell]{\sum_{i=1}^{\ell}\log \frac{P_1(Y_i)}{P_0(Y_i)}-\ell\avgD{P_1}{P_0}\leq-\nu \ell \avgD{P_1}{P_0}}\nonumber\\
    &\leq A e^{-a\ell}\text{ for some constants $A,a>0$}\\
    &\leq A e^{-a(1-\mu)\omega_n\sqrt{n}},\label{eq:reliability_concentration_4}
  \end{align}
  where the constants $A$ and $a$ are obtained using a concentration inequality, such as Hoeffding's inequality.\footnote{$\log \frac{P_1(Y)}{P_0(Y)}$ is bounded under our assumptions.} Combining, (\ref{eq:reliability_concentration_1})-(\ref{eq:reliability_concentration_4}) with~(\ref{eq:bound_iid_sum}), and substituting in~(\ref{eq:bound_pe_one_shot}), we obtain
  \begin{align}
    \E{P_{\text{err}}}\leq A e^{-a(1-\mu)\omega_n\sqrt{n}}+ e^{-\frac{1}{2}\mu^2\omega_n\sqrt{n}} + M e^{-\gamma}\left(1+e^{\omega_n^2(\zeta-1)}\right).
  \end{align}
  Hence, if
  \begin{align}
  \log  M=(1-\delta) (1-\mu)(1-\nu)\omega_n\sqrt{n}\avgD{P_1}{P_0}\text{ with $\delta\in]0;1[$},\label{eq:keyless-condition-reliability}
  \end{align}
we obtain
  \begin{align}
  \E{P_{\text{err}}}\leq A e^{-a(1-\mu)\omega_n\sqrt{n}}+ e^{-\frac{1}{2}\mu^2\omega_n\sqrt{n}} + e^{-\delta (1-\mu)(1-\nu) \omega_n\sqrt{n}\avgD{P_1}{P_0}}\left(1+e^{\omega_n^2(\zeta-1)}\right)
  \end{align}
  For $n$ large enough, with the choice of $\omega_n$ in~(\ref{eq:choice_omega}), $1+\exp\left(\omega_n^2(\zeta-1)\right)\leq e$  so that
  \begin{align}
    \E{P_{\text{err}}} \leq e^{-\rho_1 \omega_n\sqrt{n}}\quad\text{for some appropriate choice of $\rho_1>0.$}\label{eq:keyless-bound_pe}
  \end{align}

\paragraph{Channel resolvability analysis} The objective is to show that the distribution
\begin{align*}
\hQ^n(\mathbf{z})\eqdef\sum_{i=1}^M\sum_{j=1}^KW_{Z|X}^\pn(\mathbf{z}|\mathbf{x}_{ij})\frac{1}{MK}
\end{align*}
induced by the codebooks is close in divergence to $Q_{\alpha_n}^\pn(\mathbf{z})$. The proof largely follows that of\cite{Hou2014}, with the appropriate modifications. As shown in Appendix~\ref{sec:proof-lemma-resovlability-lpd-div} the divergence $\avgD{\hQ^n}{Q_{\alpha_n}^\pn}$ averaged over the random codebook satisfies the following.
\begin{lemma}
  \label{lm:lemma-resolvability-lpd-div}
  For any $\tau>0$ and for $n$ large enough,
  \begin{align}
    \E{\avgD{\hQ^n}{Q_{\alpha_n}^\pn}} \leq  n\log\frac{4}{\mu_0}\P[W_{Z|X}^\pn\Pi_{\alpha_n}^\pn]{\log \frac{W_{Z|X}^\pn(\mathbf{Z}|\mathbf{X})}{Q_0^\pn(\mathbf{Z})}\geq \tau}+\frac{e^{\tau}}{MK}.
  \end{align}
\end{lemma}
Note that
\begin{align}
\P[W_{Z|X}^\pn\Pi_{\alpha_n}^\pn]{\log \frac{W_{Z|X}^\pn(\mathbf{Z}|\mathbf{X})}{Q_0^\pn(\mathbf{Z})}\geq\tau}&=\P[W_{Z|X}^\pn\Pi_{\alpha_n}^\pn]{\sum_{i=1}^n\log \frac{W_{Z|X}({Z_i}|{X_i})}{Q_0({Z_i})}\geq\tau}.\label{eq:resolvability_analysis_beginning}
\end{align}
As in the channel reliability analysis, if $X_i=x_0$, note that $Z_i$ is distributed according to $Q_0$ and that $\log \frac{W_{Z|X}(Z_i|x_0)}{Q_0(Z_i)}=0$; if $X_i=x_1$, then $Z_i$ is distributed according to $Q_1$ and $\log \frac{W_{Z|X}(Z_i|x_1)}{Q_0(Z_i)}=\log \frac{Q_1(Z_i)}{Q_0(Z_i)}$. Hence, we may proceed as earlier, by introducing $L\eqdef \sum_{i=1}^n\mathbf{1}\{X_i=x_1\}$ and defining
\begin{align}
  \tau\eqdef(1+\mu)(1+\nu)\omega_n\sqrt{n}\avgD{Q_1}{Q_0}\quad\text{and}\quad\calD_\mu^n\eqdef\{\ell\in\mathbb{N}^*:\abs{\ell-\omega_n\sqrt{n}}<\mu\omega_n\sqrt{n}\}.
\end{align}
Intuitively, $\exp\tau$ represents the total number of codewords while $\calD_\mu$ is their likely support size. The set $\calD_\mu$ differs from $\calC_\mu$ by requiring a double-sided bound, which captures the idea that the support of codewords should not be too small for reliability but not too high either to remain covert.
Then,
\begin{align}
  \P[W_{Z|X}^\pn\Pi_{\alpha_n}^\pn]{\sum_{i=1}^n\log \frac{W_{Z|X}({Z_i}|{X_i})}{Q_0({Z_i})}\geq\tau}&\leq \sum_{\ell\in \calD^n_\mu}\P{L=\ell}\P[Q_1^\pell]{\sum_{i=1}^{\ell}\log \frac{Q_1(Z_i)}{Q_0(Z_i)}\geq\tau} + \P{L\notin \calD_\mu^n}
\end{align}
with, using a Chernoff bound,
\begin{align}
   \P{L\notin \calD_\mu^n} = \P{\abs{L-\omega_n\sqrt{n}}\leq \mu \omega_n\sqrt{n}}\leq 2e^{-\frac{1}{3}\mu^2\omega_n\sqrt{n}}.
\end{align}
For $\ell\in \calD^n_\mu$, note that
\begin{align}
 (1+\mu)(1+\nu)\omega_n\sqrt{n} -\ell >(1+\nu) \ell-\ell = \nu \ell,
\end{align}
so that
\begin{align}
  \P[Q_1^\pell]{\sum_{i=1}^{\ell}\log \frac{Q_1(Z_i)}{Q_0(Z_i)}\geq\tau} &=\P[Q_1^\pell]{{\sum_{i=1}^{\ell}\log \frac{Q_1(Z_i)}{Q_0(Z_i)}-\ell\avgD{Q_1}{Q_0}\geq\tau -\ell\avgD{Q_1}{Q_0}}}\\
  &\leq \P[Q_1^\pell]{{\sum_{i=1}^{\ell}\log \frac{Q_1(Z_i)}{Q_0(Z_i)}-\ell\avgD{Q_1}{Q_0}\geq\nu \avgD{Q_1}{Q_0}\ell}}\displaybreak[0]\\
  & \leq B e^{-b  \ell}\text{ for some constants $B,b>0$}\\
  &\leq B e^{-b (1-\mu)\omega_n\sqrt{n}},\label{eq:resolvability_analysis_end}
\end{align}
where the constants $B$ and $b$ are again obtained using Hoeffding's inequality. Combining, the inequalities~(\ref{eq:resolvability_analysis_beginning})-(\ref{eq:resolvability_analysis_end}) with Lemma~\ref{lm:lemma-resolvability-lpd-div} and choosing
\begin{align}
 \log  M+ \log  K = (1+\delta)(1+\mu)(1+\nu)\omega_n\sqrt{n}\avgD{Q_1}{Q_0},\label{eq:keyless-condition-resolvabilty}
\end{align}
we obtain 
\begin{align}
   \E{\avgD{\hQ^n}{Q_{\alpha_n}^\pn}} \leq n\log\frac{4}{\mu_0}\left(Be^{-b (1-\mu)\omega_n\sqrt{n}} +2e^{-\frac{1}{3}\mu^2\omega_n\sqrt{n}}\right)+e^{-\delta(1+\mu)(1+\nu)\omega_n\sqrt{n}\avgD{Q_1}{Q_0}}.
\end{align}
Hence, for $n$ large enough,
\begin{align}
   \E{\avgD{\hQ^n}{Q_{\alpha_n}^\pn}} \leq e^{-\rho_2\omega_n\sqrt{n}}\text{ for some appropriate choice of $\rho_2>0$}.
\end{align}

\paragraph{Identification of specific code} Choosing $\mu,\nu,\delta$, $\log  M$, and $\log  K$, to satisfy both~(\ref{eq:keyless-condition-reliability}) and~(\ref{eq:keyless-condition-resolvabilty}), Markov's inequality allows us to conclude that there exists at least one specific coding scheme with $n$ large enough and appropriate constants $\xi_1,\rho_3>0$ such that
\begin{align}
  P_{\text{err}} \leq e^{-\xi_1\omega_n\sqrt{n}}\quad\text{and}\quad \avgD{\hQ^n}{Q_{\alpha_n}^\pn}\leq e^{-\rho_3 \omega_n\sqrt{n}}.\label{eq:performance_specific_code}
\end{align}
In particular, Pinsker's inequality also ensures that $\V{\hQ^n,Q_{\alpha_n}^\pn}\leq e^{-\frac{1}{2}\rho_3\omega_n\sqrt{n}}$. Next, notice that
\begin{align}
  \avgD{\hQ^n}{Q_{0}^\pn} &= \avgD{\hQ^n}{Q_{\alpha_n}^\pn} + \sum_{\mathbf{z}}\hQ^n(\mathbf{z})\log\frac{Q_{\alpha_n}^\pn(\mathbf{z})}{Q_{0}^\pn(\mathbf{z})}\\
&=  \avgD{\hQ^n}{Q_{\alpha_n}^\pn} + \avgD{Q_{\alpha_n}^\pn}{Q_{0}^\pn}+\sum_{\mathbf{z}}\left(\hQ^n(\mathbf{z})-Q_{\alpha_n}^\pn(\mathbf{z})\right)\log\frac{Q_{\alpha_n}^\pn(\mathbf{z})}{Q_{0}^\pn(\mathbf{z})}\label{eq:bound_div_var}
\end{align}
and
\begin{align}
\abs{\sum_{\mathbf{z}}\left(\hQ^n(\mathbf{z})-Q_{\alpha_n}^\pn(\mathbf{z})\right)\log\frac{Q_{\alpha_n}^\pn(\mathbf{z})}{Q_{0}^\pn(\mathbf{z})}}\leq 2n\V{\hQ^n,Q_{\alpha_n}^\pn}\log\frac{1}{\mu_0}\leq 2ne^{-\frac{1}{2}\rho_3\omega_n\sqrt{n}}\log\frac{1}{\mu_0}.\label{eq:bound_var}
\end{align}
Hence, combining~(\ref{eq:performance_specific_code})-(\ref{eq:bound_var}), we conclude that there exists a constant $\xi_2>0$ such that, for $n$ large enough,  
\begin{align*}
  \abs{\avgD{\hQ^n}{Q_{0}^\pn}-\avgD{Q_{\alpha_n}^\pn}{Q_{0}^\pn}} \leq e^{-\xi_2\omega_n\sqrt{n}}.
\end{align*}
The statement of the theorem is finally obtained by setting $\xi\eqdef \frac{1}{2}\left((1+\delta)(1+\mu)(1+\nu)-(1-\delta)(1-\mu)(1-\nu)\right)$.
\end{IEEEproof}
\smallskip
\begin{remark}
  A closer inspection of Appendix~\ref{sec:proof-lemma-reliability-lpd} shows that Lemma~\ref{lm:reliability-lpd} applies to continuous channels. The concentration result follows with any condition that guarantees a concentration result for the sum of $n$ \ac{iid} realizations of $\log  \frac{P_1(Y)}{P_0(Y)}$. In particular, the concentration follows directly if $\log  \frac{P_1(Y)}{P_0(Y)}$ is sub-Gaussian~\cite{ConcentrationInequalities}. The adaptation of Lemma~\ref{lm:lemma-resolvability-lpd-div} to continuous channels is discussed in Section~\ref{sec:continuous-channels}.
\end{remark}

As in Section~\ref{sec:covert-comm-with}, one may also characterize the asymptotic scaling of $\log M$ and $\log K$ for the proposed scheme. We shall see in Section~\ref{sec:converse-results} that the scalings of the message and key size are optimal.
\begin{corollary}
  \label{cor:scaling-conditions}
    Consider a discrete memoryless covert communication channel with $P_1\ll P_0$, $Q_1\ll Q_0$, and $Q_1\neq Q_0$. For any $\xi\in]0;1[$, there exist covert communication schemes such that
  \begin{align*}
    \lim_{n\rightarrow\infty}&\avgD{\hQ^n}{Q_0^\pn}=0, \quad\lim_{n\rightarrow\infty}P_{\text{err}}=0,\\
    \lim_{n\rightarrow\infty}&\frac{\log M}{\sqrt{n\avgD{{\hQ}^n}{Q_0^\pn}}} = (1-\xi)  \sqrt{\frac{2}{\chid{Q_1}{Q_0}}}\avgD{P_1}{P_0},\\
    \lim_{n\rightarrow\infty}&\frac{\log K}{\sqrt{n\avgD{{\hQ}^n}{Q_0^\pn}}} = \sqrt{\frac{2}{\chid{Q_1}{Q_0}}}\left[{(1+\xi)\avgD{Q_1}{Q_0}-(1-\xi)\avgD{P_1}{P_0}}\right]^+.
  \end{align*}
\end{corollary}
\begin{IEEEproof}
The result follows as in the proof of Corollary~\ref{cor:scaling-key-assisted} and is omitted for brevity.
\end{IEEEproof}

\section{Converse result for \acp{DMC}}
\label{sec:converse-results}
In this section, we show the optimality of the asymptotic limits given in Corollary~\ref{cor:scaling-conditions}. The proof leverages the converse technique and results of~\cite{Wang2015,Wang2015a,Hou2014a}.
\begin{theorem}
  \label{th:converse}
      Consider a discrete memoryless covert communication channel with $P_1\ll P_0$, $Q_1\ll Q_0$, and $Q_1\neq Q_0$. Consider a sequence of covert communication schemes with increasing blocklength $n$ characterized by $\epsilon_n\eqdef P_{\text{err}}$ and $\delta_n\eqdef \avgD{\hQ^n}{Q_0^\pn}$.  If $\lim_{n\rightarrow\infty}\epsilon_n=\lim_{n\rightarrow\infty}\delta_n=0$, we have
  \begin{align}
    \lim_{n\rightarrow\infty}\frac{\log  M}{\sqrt{n \avgD{{\hQ}^n}{Q_0^\pn}}} &\leq \sqrt{\frac{2}{\chid{Q_1}{Q_0}}}\avgD{P_1}{P_0}\label{eq:converse_1}.
  \end{align}
For a sequence of schemes such that~(\ref{eq:converse_1}) holds with equality, we have
\begin{align}
      \lim_{n\rightarrow\infty}\frac{\log  M+\log  K}{\sqrt{n \avgD{{\hQ}^n}{Q_0^\pn}}} &\geq \sqrt{\frac{2}{\chid{Q_1}{Q_0}}}\avgD{Q_1}{Q_0}.\label{eq:converse_2}
\end{align}
\end{theorem}
\begin{IEEEproof}
  The proof of~(\ref{eq:converse_1}) is an adaptation of~\cite[Proof of Theorem 2]{Wang2015a}. The proof of~(\ref{eq:converse_2}) follows by adapting the steps of~\cite[Section 5.2.3]{Hou2014a} to lower bound the sum $\log  M+\log  K$. We detail here the modifications required to analyze the present setting.

Consider a sequence of length-$n$ codes for the setting in Fig.~\ref{fig:covert_comm} with $\epsilon_n\eqdef P_{\text{err}}$ and $\delta_n\eqdef\avgD{\smash{\widehat{Q}^n}}{Q_0^\pn}$, such that 
$\lim_{n\rightarrow\infty}\epsilon_n=  \lim_{n\rightarrow\infty}\delta_n=0$, and $\log M$ takes the maximum value such that $\lim_{n\rightarrow\infty}\log M =\infty$. We start by upper bounding $\log M$ using standard techniques.
\begin{align}
  \log M = \avgH{W} &= \avgI{W;Y^nS}+\avgH{W|Y^nS}\\
  &\leq \avgI{W;Y^nS} + \Hb{\epsilon_n}+\epsilon_n\log M\\
  &= \avgI{W;Y^n|S} + \Hb{\epsilon_n}+\epsilon_n\log M\\
  &\leq \avgI{WS;Y^n} + \Hb{\epsilon_n}+\epsilon_n\log M\\
                    &\leq \avgI{X^n;Y^n} + \Hb{\epsilon_n}+\epsilon_n\log M\\
&\leq n\avgI{\smash{\tilde{X};\tilde{Y}}}+\Hb{\epsilon_n}+\epsilon_n\log M,\label{eq:upper_bound_logM}
\end{align}
where the random variables $\tilde{X}$ and $\tilde{Y}$ have distribution
\begin{align}
  P_{\tilde{X}}(x)\eqdef \frac{1}{n}\sum_{i=1}^nP_{X_i}(x)\quad\text{and}\quad   P_{\tilde{X}\tilde{Y}}(x,y)\eqdef   P_{\tilde{X}}(x)W_{Y|X}(y|x).
\end{align}
Hence,
\begin{align}
\log M \leq \frac{1}{1-\epsilon_n}\left(n\avgI{\smash{\tilde{X};\tilde{Y}}} + \Hb{\epsilon_n}\right).  \label{eq:converse_bound_logM_reliability}
\end{align}
Following~\cite{Hou2014a}, we obtain
\begin{align}
  \log M + \log K &\geq \avgH{W S}\\
  &\geq \avgI{WS;Z^n}\\
  &\stackrel{(a)}{\geq} \avgI{X^n;Z^n}\\
  &\stackrel{(b)}{\geq} n \avgI{\smash{\tilde{X};\tilde{Z}}}-\delta_n,\label{eq:lower_bound_logMK}
\end{align}
where $(a)$ follows because $X^n$ is a function of $W$ and $S$, and $(b)$ follows by the steps of~\cite[Section 5.2.3]{Hou2014a} upon defining the random variables $\tilde{X}$ and $\tilde{Z}$ to have joint distribution
\begin{align}
  P_{\tilde{X}\tilde{Z}}(x,z)\eqdef   P_{\tilde{X}}(x)W_{Z|X}(z|x).
\end{align}
Following the reasoning of~\cite{Wang2015a,Hou2014a}, one can show
\begin{align}
  \delta_n=\avgD{\hQ^n}{Q_0^\pn} \geq n \avgD{\tQ}{Q_0}\quad\text{with }\tQ(z)\eqdef\sum_{i=1}^n\frac{1}{n}\hQ_i(z).\label{eq:lower_bound_converse_div}
\end{align}
Applying Pinsker's inequality, we see that $\lim_{n\rightarrow\infty}\V{\tQ,Q_0}=0$ so that $\forall z$ $\lim_{n\rightarrow\infty}\tQ(z)=Q_0(z)$, and $P_{\tilde{X}}$ must be of the form
\begin{align*}
  P_{\tilde{X}}(x) = (1-\mu_n) \mathbf{1}\left\{x=x_0\right\} + \mu_n\mathbf{1}\left\{x=x_1\right\}\quad\text{with }\lim_{n\rightarrow\infty}\mu_n=0.
\end{align*}
Using the notation of Section~\ref{sec:covert-comm-over}, we may write $P_{\tilde{X}}=\Pi_{\mu_n}$, $\tQ=Q_{\mu_n}$, and $P_{\tilde{Y}}=P_{\mu_n}$. Using the bounds given in Lemma~\ref{lm:preliminary}, we find that
\begin{align}
 \frac{\mu_n^2}{2}\chid{Q_1}{Q_0}\left(1-\sqrt{\mu_n}\right)\leq \avgD{Q_{\mu_n}}{Q_0} &\leq \frac{\mu_n^2}{2}\chid{Q_1}{Q_0}\left(1+\sqrt{\mu_n}\right),\label{eq:bound_converse_D}\\
  \avgI{\smash{\tilde{X};\tilde{Y}}} &\leq \mu_n\avgD{P_1}{P_0}-\avgD{P_{\mu_n}}{P_0}\leq \mu_n\avgD{P_1}{P_0},\label{eq:bound_converse_IXY}\\
  \avgI{\smash{\tilde{X};\tilde{Z}}}&=  \mu_n\avgD{Q_1}{Q_0}-\avgD{Q_{\mu_n}}{Q_0}. \label{eq:bound_converse_IXZ}
\end{align}
Note that the lower bound of $\avgD{Q_{\mu_n}}{Q_0}$ in~(\ref{eq:bound_converse_D}) combined with the inequality in~(\ref{eq:lower_bound_converse_div}) imposes that $\lim_{n\rightarrow\infty}\sqrt{n}\mu_n=0$. The constraint $\lim_{n\rightarrow\infty}\log M =\infty$ combined with~(\ref{eq:converse_bound_logM_reliability}) and~(\ref{eq:bound_converse_IXY}) also requires that $\lim_{n\rightarrow\infty} n\mu_n =\infty$. Hence,
\begin{align*}
  \frac{\log M}{\sqrt{n \avgD{{\hQ}^n}{Q_0^\pn}}} &\leq \frac{n\avgI{\smash{\tilde{X};\tilde{Y}}}+\Hb{\epsilon_n}}{(1-\epsilon_n)\sqrt{n^2\avgD{Q_{\mu_n}}{Q_0}}}\displaybreak[0]\\
                                                  &\leq \frac{\mu_n\avgD{P_1}{P_0}+\frac{1}{n}\Hb{\epsilon_n}}{(1-\epsilon_n)\sqrt{\frac{1}{2}\mu_n^2 \chid{Q_1}{Q_0}(1-\sqrt{\mu_n})}}\displaybreak[0]\\
                                                  &\leq  \sqrt{\frac{2}{\chid{Q_1}{Q_0}}}\frac{ \avgD{P_1}{P_0}+\frac{1}{n\mu_n}}{(1-\epsilon_n)\sqrt{\left(1-\sqrt{\mu_n}\right)}},
\end{align*}
and
\begin{align}
      \lim_{n\rightarrow\infty}  \frac{\log M}{\sqrt{n \avgD{{\hQ}^n}{Q_0^n}}}\leq  \sqrt{\frac{2}{\chid{Q_1}{Q_0}}}\avgD{P_1}{P_0}.\label{eq:upper-bound-logM}
\end{align}
For any sequence of codes such that~(\ref{eq:upper-bound-logM}) holds with equality, which we know is indeed possible from Corollary~\ref{cor:scaling-conditions},~\eqref{eq:converse_bound_logM_reliability} combined with~(\ref{eq:bound_converse_IXY}) and  $\lim_{n\rightarrow\infty} n\mu_n =\infty$ impose  that for any $\rho>0$,
\begin{align}
  (1-\rho)\sqrt{\frac{2}{\chid{Q_1}{Q_0}}}\avgD{P_1}{P_0}\leq \lim_{n\rightarrow\infty}  \frac{n\mu_n\avgD{P_1}{P_0}}{\sqrt{n \avgD{{\hQ}^n}{Q_0^n}}}.\label{eq:lower_bound_mus}
\end{align}
Hence, combing~(\ref{eq:lower_bound_logMK}),~(\ref{eq:bound_converse_IXZ}),~(\ref{eq:lower_bound_mus}), we obtain for any $\rho>0$
\begin{align*}
    \frac{\log M+\log K}{\sqrt{n \avgD{{\hQ}^n}{Q_0^\pn}}} &\geq \frac{n\mu_n \avgD{Q_1}{Q_0}-n\avgD{Q_{\mu_n}}{Q_0}-\delta_n}{\sqrt{n \avgD{{\hQ}^n}{Q_0^\pn}}}\\
  &\geq   (1-\rho)\sqrt{\frac{2}{\chid{Q_1}{Q_0}}}\left(\avgD{Q_1}{Q_0}-\frac{1}{\mu_n}\avgD{Q_{\mu_n}}{Q_0}-\frac{\delta_n}{n\mu_n}\right).
\end{align*}
Since $\rho>0$ is arbitrary, we have
\begin{align*}
\lim_{n\rightarrow\infty}\frac{\log M+\log K}{\sqrt{n \avgD{{\hQ}^n}{Q_0^\pn}}} \geq \sqrt{\frac{2}{\chid{Q_1}{Q_0}}}\avgD{Q_1}{Q_0}.
\end{align*} 
\end{IEEEproof}
If $W_{Z|X}=W_{Y|X}$, the right hand side of~(\ref{eq:converse_1}) is actually a special case of~\cite[Theorem 2]{Wang2015a} for two inputs.

\section{Extensions and applications}
\label{sec:extens-appl}

\subsection{Non-vanishing $\avgD{\smash{\widehat{Q}^n}}{\smash{Q_0^{\otimes n}}}$}
\label{sec:non-vanishing}

Instead of requiring that $\lim_{n\rightarrow\infty}\avgD{\hQ^n}{Q_0^\pn}=0$, we could relax the constraint by asking that $\lim_{n\rightarrow\infty}\avgD{\hQ^n}{Q_0^\pn}=\delta$ for some chosen $\delta>0$. The optimal scalings of $\log M$ and $\log K$ with $n$ obtained in this case are summarized in Table~\ref{tab:summary_scaling_non_vanish} and Table~\ref{tab:summary_scaling_key_non_vanish}, respectively. The result when $Q_1\ll Q_0$ and $P_1\ll P_0$ is obtained by choosing a sequence $\omega_n$ such that  $\lim_{n\rightarrow\infty}\omega_n=\omega_0>0$ in the proof of Theorem~\ref{th:key-assist-2}. The results for the other situations are obtained with the same modification in the analysis of Appendix~\ref{sec:spec-cases-chann}.

\begin{table}[h]
  \centering
  \caption{Optimal scaling of $\log M$ for which $\lim_{n\rightarrow\infty}\avgD{\smash{\widehat{Q}^n}}{\smash{Q_0^{\otimes n}}}=\delta>0$.}
\smallskip
  \begin{tabular}{|>{$}c<{$}|>{$}c<{$}>{$}c<{$}>{$}c<{$}|}
    \hline    &P_1\ll P_0& P_1 \nll P_0 & P_1 = P_0\\\hline
    Q_1\ll Q_0    &  \Theta(\sqrt{n})&\Theta(\sqrt{n}\log n)&0\\
    Q_1\nll Q_0    &0&0&0\\
    Q_1=Q_0    &\Theta(n)&\Theta(n)&0\\\hline
  \end{tabular}
  \label{tab:summary_scaling_non_vanish}
\end{table}

\begin{table}[h]
  \centering
  \caption{Optimal scaling of $\log K$ for which $\lim_{n\rightarrow\infty}\avgD{\smash{\widehat{Q}^n}}{\smash{Q_0^{\otimes n}}}=\delta>0$.}
\smallskip
  \begin{tabular}{|>{$}c<{$}|>{$}c<{$}>{$}c<{$}>{$}c<{$}|}
    \hline    &P_1\ll P_0& P_1 \nll P_0 & P_1 = P_0\\\hline
    Q_1\ll Q_0    &  \Theta(\sqrt{n})&0&0\\
    Q_1\nll Q_0    &0&0&0\\
    Q_1=Q_0    &0&0&0\\\hline
  \end{tabular}
  \label{tab:summary_scaling_key_non_vanish}
\end{table}

\subsection{Multiple symbols}
\label{sec:multiple-symbols}
A close inspection of the proofs shows that the calculations may be extended to multiple symbols $\{x_i\}_{i\in\intseq{1}{N}}$ such that $\forall i\in\intseq{1}{N}\;x_i\neq x_0$. Specifically, assume that each symbol $x_i$ is assigned probability $p_i\alpha_n$, with $\sum_{i=1}^Np_i=1$. Denote $P_i\eqdef W_{Y|X=x_i}$ and $Q_i\eqdef W_{Z|X=x_i}$. Following verbatim the approach of Section~\ref{sec:stealth-processes}, one may redefine 
\begin{align}
  Q_{\alpha_n}(z)=\alpha_n\sum_{i=1}^N p_iQ_i(z)+(1-\alpha_n)Q_0(z),
\end{align}
so that for all $n\in\mathbb{N}^*$,
\begin{align}
\avgD{Q_{\alpha_n}}{Q_0}\leq   \frac{\alpha_n^2}{2}\chid{\sum_i p_i Q_i}{Q_0} -\frac{\alpha_n^3}{6}     \chid[3]{\sum_i p_i Q_i}{Q_0} +\frac{\alpha_n^4}{3}    \chid[4]{\sum_i p_i Q_i}{Q_0}.\label{eq:upper_bound_d_covert_multi}
\end{align}
and for $n$ large enough, 
\begin{multline}
  \avgD{Q_{\alpha_n}}{Q_0} \geq
\frac{\alpha_n^2}{2}\chid{\sum_i p_i Q_i}{Q_0} -\alpha_n^3\left(\frac{1}{2}\chid[3]{\sum_i p_i Q_i}{Q_0} -\frac{1}{3}\etad[3]{\sum_i p_i Q_i}{Q_0}\right)\\
+\frac{2\alpha_n^4}{3}\etad[4]{\sum_i p_i Q_i}{Q_0}. \label{eq:lower_bound_d_covert_multi}
\end{multline}
Following the proof of Theorem~\ref{th:key-assist-2}, we then obtain the following result.
\begin{theorem}
  \label{th:general-result}
  Consider a discrete memoryles covert communication channel such that $Q_0$ is is not a mixture of $\{Q_i\}_{i\in\intseq{1}{N}}$ and $\forall i\in\intseq{1}{N}$ $Q_i\ll Q_0$ and $P_i\ll P_0$. Let $\{p_i\}_{i\in\intseq{1}{N}}\in[0;1]^N$ be such that $\sum_{i=1}^N p_i =1$ and let $\alpha_n\eqdef\frac{\omega_n}{\sqrt{n}}$ with $\omega_n\in o(1)\cap\omega(\frac{1}{\sqrt{n}})$ as $n\rightarrow\infty$. For any $\xi\in]0;1[$, there exist $\xi_1, \xi_2>0$ depending on $\xi$, $W_{Y|X}$, $W_{Z|X}$, and a covert communication scheme as in Fig.~\ref{fig:channel-resol-covert} such that, for $n$ large enough,
  \begin{align*}
    \log M &= (1-\xi)\omega_n\sqrt{n}\left(\sum_i p_i\avgD{P_i}{P_0}\right)\\
    \log K &= \omega_n\sqrt{n}\left[ (1+\xi) \sum_{i=1}^N p_i\avgD{Q_i}{Q_0}-(1-\xi)\sum_ip_i\avgD{P_i}{P_0}\right]^+
  \end{align*}
  and
\begin{align*}
  P_{\text{err}}\leq e^{-\xi_1\omega_n\sqrt{n}},\qquad    \abs{\avgD{\hQ^n}{Q_{0}^\pn}-\avgD{Q_{\alpha_n}^\pn}{Q_{0}^\pn}} \leq e^{-\xi_2\omega_n\sqrt{n}},
  \end{align*}
\end{theorem}
In particular, we obtain a characterization of the asymptotic scaling.
\begin{corollary}
\label{cor:general_scaling}
Consider a discrete memoryles covert communication channel such that $Q_0$ is is not a mixture of $\{Q_i\}_{i\in\intseq{1}{N}}$ and $\forall i\in\intseq{1}{N}$ $Q_i\ll Q_0$ and $P_i\ll P_0$. Let $\{p_i\}_{i\in\intseq{1}{N}}\in[0;1]^N$ be such that $\sum_{i=1}^N p_i =1$. Then, there exist covert communication schemes such that
  \begin{align*}
    \lim_{n\rightarrow\infty}&\avgD{\hQ^n}{Q_0^\pn}=0, \quad    \lim_{n\rightarrow\infty}P_{\text{err}}=0,\\
    \lim_{n\rightarrow\infty}&\frac{\log M}{\sqrt{n\avgD{{\hQ}^n}{Q_0^\pn}}} =  \sqrt{\frac{2}{\chid{\sum_{i=1}^N p_i Q_i}{Q_0}}}\sum_{i=1}^N p_i \avgD{P_i}{P_0},\\
    \lim_{n\rightarrow\infty}&\frac{\log K}{\sqrt{n\avgD{{\hQ}^n}{Q_0^\pn}}} = \sqrt{\frac{2}{\chid{\sum_{i=1}^N p_i Q_i}{Q_0}}}\left[\sum_{i=1}^N p_i \left({\avgD{Q_i}{Q_0}-\avgD{P_i}{P_0}}\right)\right]^+.
  \end{align*}
\end{corollary}
One can also show the optimality of the scalings by following~\cite{Wang2015,Wang2015a} and adapting the proof of Theorem~\ref{th:converse}.

\subsection{Covert and secret communication}
\label{sec:covert-secr-co mm}
The problem as formulated in Section~\ref{sec:covert-comm-over} only requires communication to be undetectable but does not prevent the warden from extracting information about the transmitted message. To address this, one could consider an additional semantic secrecy~\cite{Bellare2012} constraint of the form
\begin{align}
  \forall p_W\,\lim_{n\rightarrow\infty}\avgI{W;Z^n} = 0.\label{eq:semantic_secrecy}
\end{align}
The problem is then similar to the effective secrecy introduced in~\cite{Hou2014,Han2014} in a regime of undetectable communication, and similar to the ``hidable and deniable'' communication setting in~\cite{Che2014}. 

The architecture studied in Section~\ref{sec:covert-comm-with} already satisfies this condition because the modulation as per~(\ref{eq:modulation}) performs a one-time pad of the encoded message bits with the key bits $\widehat{S}$. If one does not wish to use an extra key for secrecy, then the next theorem shows that semantic secrecy may be obtained ``for free''  when $\avgD{P_1}{P_0}>\avgD{Q_1}{Q_0}$, by using a code for the wiretap channel instead of a code for reliable communication.

\begin{theorem}
  \label{th:secrecy}
    Consider a discrete memoryless covert communication channel with $P_1\ll P_0$, $Q_1\ll Q_0$, and $Q_1\neq Q_0$. Let $\alpha_n\eqdef\frac{\omega_n}{\sqrt{n}}$ with $\omega_n\in o(1)\cap\omega(\frac{1}{\sqrt{n}})$ as $n\rightarrow\infty$. For any $\xi\in]0;1[$, there exist $\xi_1, \xi_2, \xi_3>0$ depending on $\xi$, $W_{Y|X}$, $W_{Z|X}$, and a covert communication scheme such that, for $n$ large enough,
  \begin{align*}
    \log M &= (1-\xi)\omega_n\sqrt{n}\avgD{P_1}{P_0},\quad    \log K = (1+\xi)\omega_n\sqrt{n}\avgD{Q_1}{Q_0},
  \end{align*}
and
\begin{align*}
P_{\text{err}}\leq e^{-\xi_1\omega_n\sqrt{n}},\quad \abs{\avgD{\hQ^n}{Q_{0}^\pn}-\avgD{Q_{\alpha_n}^\pn}{Q_{0}^\pn}} \leq e^{-\xi_2\omega_n\sqrt{n}}, \quad\forall p_W\, \avgI{W;Z^n} \leq e^{-\xi_3\omega_n\sqrt{n}}.
\end{align*}
\end{theorem}
\begin{IEEEproof}
  We only sketch the details here for brevity. Let $\xi\in]0;1[$ and $n\in\mathbb{N}^*$ sufficiently large. 

  If $(1-\xi)\avgD{P_1}{P_0}\leq(1+\xi)\avgD{Q_1}{Q_0}$, we know from Theorem~\ref{th:key-assist-2} that we may transmit $\log M=(1-\xi)\omega_n\sqrt{n}\avgD{P_1}{P_0}$ message bits with the help of $\omega_n\sqrt{n}\left((1+\xi)\avgD{Q_1}{Q_0}-(1-\xi)\avgD{P_1}{P_0}\right)$ key bits. One may render the message bits secret by performing a one-time pad requiring another $(1-\xi)\omega_n\sqrt{n}\avgD{P_1}{P_0}$ key bits, for a total of $(1+\xi)\omega_n\sqrt{n}\avgD{Q_1}{Q_0}$ key bits.

  If $(1-\xi)\avgD{P_1}{P_0}>(1+\xi)\avgD{Q_1}{Q_0}$, we modify the random coding argument of Theorem~\ref{th:key-assist-2} as follows. Let $M,M'\in\mathbb{N}^*$. Generate $MM'$ codewords $\mathbf{x}_{ij}\in\{x_0,x_1\}^n$ with $i\in\intseq{1}{M}$ and $j\in\intseq{1}{M'}$. The index $i$ is used to encode a message $W$ while $j$ is used to encode another message $W'$. Following the exact same reliability analysis as in the proof of Theorem~\ref{th:key-assist-2}, we conclude that if
  \begin{align}
    \log M+\log M' &= (1-\xi)\omega_n\sqrt{n}\avgD{P_1}{P_0}
  \end{align}
  then $\E{P_{\text{err}}}\leq e^{-\rho_1\omega_n\sqrt{n}}$ for some $\rho_1>0$. Following the principle of achieving secrecy from resolvability\cite{Bloch2011e}, we may also prove that if
  \begin{align}
    \log M' &= (1+\xi)\omega_n\sqrt{n}\avgD{Q_1}{Q_0}
  \end{align}
  then $\E{\avgI{W;Z^n}} \leq e^{-\rho_3\omega_n\sqrt{n}}$ for some $\rho_3>0$. In addition, since $\log M+\log M'\geq (1+\xi)\omega_n\sqrt{n}\avgD{Q_1}{Q_0}$, covertness follows ``for free'' using  the same arguments as in Theorem~\ref{th:key-assist-2}. Finally, the bits of $W'$ may be protected by a one-time pad, requiring $(1+\xi)\omega_n\sqrt{n}\avgD{Q_1}{Q_0}$ key bits. The expurgation argument leading to semantic secrecy is standard, e.g.,~\cite[Lemma 1]{Renes2011}. 
\end{IEEEproof}

\begin{figure}[h!]
  \centering
  \includegraphics[scale=0.85]{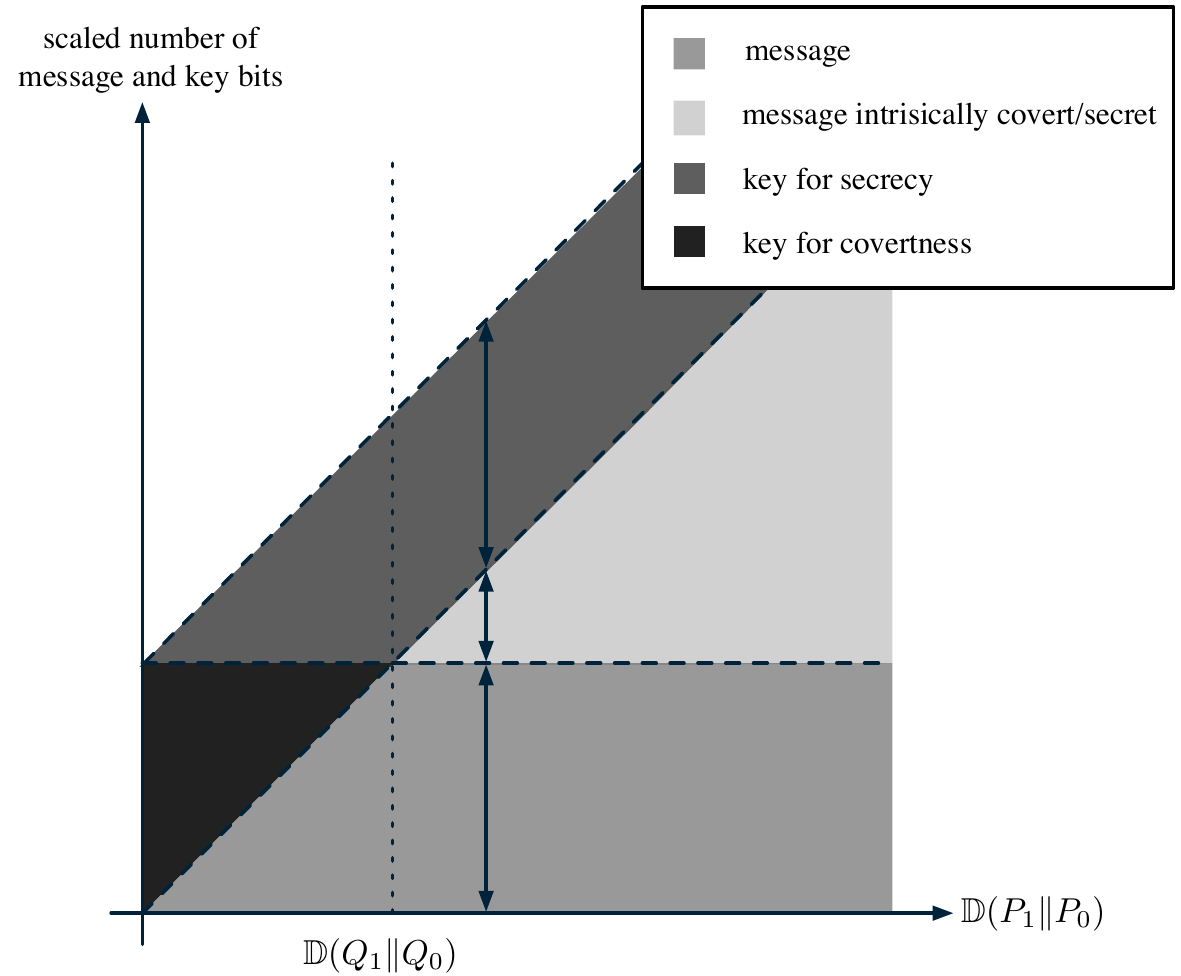}
  \caption{Illustration of different regimes of covert and secret communication.}
  \label{fig:regimes}
\end{figure}

The different regimes of covert and secret communication are illustrated in Fig.~\ref{fig:regimes}, which shows the asymptotic number of messages bits and keys bits scaled by $\sqrt{\frac{2}{\chid{Q_1}{Q_0}}}\omega_n\sqrt{n}$ as a function of $\avgD{P_1}{P_0}$ for a fixed value of $\avgD{Q_1}{Q_0}$. As depicted by the different colors in Fig.~\ref{fig:regimes}, the key bits may be used either for covertness or for secrecy. Similarly, some messages bits are intrisically covert and secret, while others require the use of a secret-key. For $\avgD{P_1}{P_0}\leq\avgD{Q_1}{Q_0}$, secret-keys are required for both covertness and secrecy while for $\avgD{P_1}{P_0}>\avgD{Q_1}{Q_0}$, secret keys are only required for added secrecy. Irrespective of the regime, the total number of secret key bits remains the same.

\subsection{Gaussian channels}
\label{sec:continuous-channels}
Gaussian channels are of particular practical interest with the innocent symbol $x_0=0$. Lemma~\ref{lm:reliability-lpd} still applies to continuous channels but Lemma~\ref{lm:lemma-resolvability-lpd-div} does not since $\mu_0=0$. Nevertheless, one may establish a slightly weaker result in terms of the total variation. Since $\alpha+\beta\geq 1-\V{\hQ^n,Q_0^\pn}$, it suffices to establish that $\V{\hQ^n,Q_0^\pn}$ vanishes to ensure covert communications. As shown in Appendix~\ref{sec:proof-lemma-resovability-lpd-var} one may adapt the proof of~\cite[Theorem VII.1]{Cuff2013} to establish the following.
\begin{lemma}
\label{lm:lemma-resolvability-lpd-var}  
For any channel $(\calX,W_{Z|X},\calZ)$ and for any $\tau>0$,
\begin{align}
    \E{\V{\hQ^n,Q_{\alpha_n}^\pn}} \leq \P[W_{Z|X}^\pn\Pi_{\alpha_n}^\pn]{\log \frac{W^\pn_{Z|X}(\mathbf{Z}|\mathbf{X})}{Q_0^\pn (\mathbf{Z})}>\tau} + \frac{1}{2}\sqrt{\frac{e^\tau}{MK}}.
\end{align}
\end{lemma}
Consequently, one may establish the following result.
\begin{theorem}
  \label{th:continuous_covert}
  Consider a continuous memoryless covert communication channel with $P_1\ll P_0$, $Q_1\ll Q_0$, and $Q_1\neq Q_0$. Assume that the random variables $\log \frac{Q_1(Z)}{Q_0(Z)}$ with $Z\sim Q_1$ and $\log \frac{P_1(Y)}{P_0(Y)}$ with $Y\sim P_1$ are sub-Gaussian, and $\int \frac{P_1(y)^2}{P_0(y)}\dd y<\infty$. let $\alpha_n\eqdef\frac{\omega_n}{\sqrt{n}}$ with $\omega_n\in o(1)\cap\omega(\frac{1}{\sqrt{n}})$ as $n\rightarrow\infty$. For any $\xi\in ]0;1[$, there exist $\xi_1, \xi_2>0$ depending on $\xi$, $W_{Y|X}$, $W_{Z|X}$, and a covert communication scheme as in Fig.~\ref{fig:channel-resol-covert} such that, for $n$ large enough,
  \begin{align*}
    \log M &= (1-\xi)\omega_n\sqrt{n}\avgD{P_1}{P_0}\\
    \log K &= \omega_n\sqrt{n}\left[ (1+\xi) \avgD{Q_1}{Q_0}-(1-\xi)\avgD{P_1}{P_0}\right]^+
  \end{align*}
  and
\begin{align*}
P_{\text{err}}\leq e^{-\xi_1\omega_n\sqrt{n}},\qquad    \V{\hQ^n,Q_{\alpha_n}^\pn}\leq e^{-\xi_2\omega_n\sqrt{n}}.
  \end{align*}  
\end{theorem}
For an \ac{AWGN} channel, note that $P_i\sim\calN(x_i,\sigma)$. One may check that $\log \frac{P_1(Y)}{P_0(Y)}$ for $Y\sim P_1$ is sub-Gaussian since
\begin{align}
  \log \frac{P_1(Y)}{P_0(Y)} = \frac{x_1}{\sigma^2}Y  -\frac{x_1^2}{2\sigma_2},
\end{align}
which is a Gaussian random variable. Also,
\begin{align}
  \int \frac{P_1(y)^2}{P_0(y)}\dd y = e^{-\frac{x_1^2}{\sigma^2}}<\infty.
\end{align}
One can finally show covertness, by using the triangle inequality to obtain
\begin{align}
  \V{\hQ^n,Q_0^\pn}&\leq   \V{\hQ^n,{Q}_{\alpha_n}^\pn}+  \V{{Q}_{\alpha_n}^\pn,Q_0^\pn} \leq e^{-\xi_2\omega_n\sqrt{n}}+ \sqrt{n \avgD{Q_{\alpha_n}}{Q_0}}.
\end{align}
As in the case of \acp{DMC}, no key is required if $\avgD{P_1}{P_0}>\avgD{Q_1}{Q_0}$.

\section{Acknowledgements}
\label{sec:aknowledgements}

The author would like to thank Boulat Bash, Sid Jaggi, Gerhard Kramer, and Ligong Wang for stimulating and enjoyable discussions. In particular, the final presentation of the present paper and the converse results largely benefited from~\cite{Wang2015a,Hou2014a}. This work was supported in part by grants NSF CIF-1320298, CNS-1527387, and the ANR-13-BS03-0008.

\newpage
\appendices

\section{\acf{KL} divergence and hypothesis testing}
\label{sec:kullb-leibl-diverg}

In this section, we provide a brief discussion in the spirit of~\cite{Cachin2004} to provide an alternative operational significance of the \ac{KL} divergence for hypothesis testing. While \ac{KL} divergence naturally appears in the exponents of the probability of false alarm and missed-detection when testing whether an \ac{iid} process is generated according to one of two different distributions, this interpretation is not valid in the present setting since the distribution $\hQ^n$ is not \ac{iid}. 

Nevertheless, assume that the warden's hypothesis test is characterized by Type I error $\alpha$, Type II error $\beta$, and a rejection region $\calR$. This hypothesis test may be viewed as a ``black box'' that outputs a Bernoulli random variable $B_p$, indicating ``0'' if $H_0$ is accepted, or ``1'' if $H_0$ is rejected. If $H_0$ is true then $p=\alpha$ by definition; alternatively, if $H_1$ is true, then $p=1-\beta$ by definition. The performance of the test may be captured by computing the Jensen-Shannon divergence~\cite{Lin1991a}
\begin{align}
  \mathbb{J}(B_\alpha,B_{1-\beta})\eqdef \avgD{B_\alpha}{B_{\frac{\alpha+1-\beta}{2}}}+\avgD{B_{1-\beta}}{B_{\frac{\alpha+1-\beta}{2}}}=\frac{1}{2}\avgD{B_{1-\beta}}{B_{\alpha}}-\avgD{B_{\frac{\alpha+1-\beta}{2}}}{B_\alpha}.
\end{align}
In fact, it is known that\footnote{with a $\log$ to the base 2} $0\leq \mathbb{J}(B_\alpha,B_{1-\beta})\leq 1$, with $\mathbb{J}(B_\alpha,B_{1-\beta})=0$ if and only if $\alpha+\beta=1$, and $\mathbb{J}(B_\alpha,B_{1-\beta})=1$ if and only if $\alpha=\beta=0$. Hence, the value of $\mathbb{J}(B_\alpha,B_{1-\beta})$ is an indication of how effective the test is. 

To achieve covert communication, one must therefore ensure that $\mathbb{J}(B_\alpha,B_{1-\beta})$ is small. By application of the log-sum inequality, one obtains
\begin{align}
2\mathbb{J}(B_\alpha,B_{1-\beta})&\leq   \avgD{B_{1-\beta}}{B_\alpha}\\
                                 &= (1-\beta)\log\frac{1-\beta}{\alpha} + \beta\log \frac{\beta}{1-\alpha}\\
                                       &=\P[\hQ^n]{\calR}\log\frac{\P[\hQ^n]{\calR}}{\P[Q_0^\pn]{\calR}} + \P[\hQ^n]{\calR^c}\log\frac{\P[\hQ^n]{\calR^c}}{\P[Q_0^\pn]{\calR^c}}\\
                                       &\leq \sum_{\mathbf{z}}\hQ^n(\mathbf{z})\log\frac{\hQ^n(\mathbf{z})}{Q_0^\pn(\mathbf{z})}\\
                                       &= \avgD{\hQ^n}{Q_0^\pn}.
\end{align}
Hence, a sufficient condition to make the test ineffective is again to minimize $\avgD{\hQ^n}{Q_0^\pn}$.

\section{Proof of Lemma~\ref{lm:preliminary}}
\label{sec:proof-lemma}
Note that
  \begin{align}
    \avgD{Q_{\alpha_n}}{Q_0}&=\sum_{z\in\calZ}Q_{\alpha_n}(z)\log  \frac{Q_{\alpha_n}(z)}{Q_0(z)}=\sum_{z\in\calZ}\left(Q_0(z)+\alpha_n(Q_1(z)-Q_0(z))\right)\log  \left(1+\alpha_n\frac{Q_1(z)-Q_0(z)}{Q_0(z)}\right).
  \end{align}
Using the inequality $\log (1+x)<x-\frac{x^2}{2}+\frac{x^3}{3}$ for $x>-1$, we obtain
\begin{align}
      \avgD{Q_{\alpha_n}}{Q_0}&\leq  \sum_{z\in\calZ}\left(Q_0(z)+\alpha_n(Q_1(z)-Q_0(z))\right)\left(\alpha_n\frac{Q_1(z)-Q_0(z)}{Q_0(z)}-\frac{\alpha_n^2}{2}\left(\frac{Q_1(z)-Q_0(z)}{Q_0(z)}\right)^2\right.\nonumber\\
    &\phantom{==================================}\left.+\frac{\alpha_n^3}{3}\left(\frac{Q_1(z)-Q_0(z)}{Q_0(z)}\right)^3\right)\\
&=\frac{\alpha_n^2}{2}\chid{Q_1}{Q_0} -\frac{\alpha_n^3}{6}\chid[3]{Q_1}{Q_0}  +\frac{\alpha_n^4}{3}\chid[4]{Q_1}{Q_0},\label{eq:upper_bound_d}
  \end{align}
Using the inequalities $\log (1+x)>x-\frac{x^2}{2}$ for $x\geq 0$ and $\log (1+x)>x-\frac{x^2}{2}+\frac{2x^3}{3}$ for $x\in[-\tfrac{1}{2};0]$, we obtain for $\alpha_n$ small enough,\footnote{$\alpha_n$ should be such that $\forall z\in\calZ$ with $Q_0(z)>0$ and $Q_1(z)-Q_0(z)<0$ we have $\alpha_n(Q_1(z)-Q_0(z))\geq -\frac{1}{2}Q_0(z)$}
\begin{align}
  \avgD{Q_{\alpha_n}}{Q_0} &\geq\sum_{z\in\calZ}\left(Q_0(z)+\alpha_n(Q_1(z)-Q_0(z))\right)\left(\alpha_n\frac{Q_1(z)-Q_0(z)}{Q_0(z)}-\frac{\alpha_n^2}{2}\left(\frac{Q_1(z)-Q_0(z)}{Q_0(z)}\right)^2\right)\nonumber\\
                      &\phantom{=========}+\sum_{z\in\calZ:Q_1(z)-Q_0(z)<0}\left(Q_0(z)+\alpha_n(Q_1(z)-Q_0(z))\right)\frac{2\alpha_n^3}{3}\left(\frac{Q_1(z)-Q_0(z)}{Q_0(z)}\right)^3\displaybreak[0]\\
                           &= \frac{\alpha_n^2}{2}\chid{Q_1}{Q_0}+\alpha_n^{3}\left(\frac{2}{3}\etad[3]{Q_1}{Q_0}-\frac{1}{2}\chid[3]{Q_1}{Q_0} \right)+\frac{2\alpha_n^4}{3}\etad[4]{Q_1}{Q_0}. \label{eq:lower_bound_d}
\end{align}
Finally, note that
  \begin{align}
    \avgI{P_X;W_{Z|X}} &= (1-\alpha_n)\avgD{Q_0}{Q_{\alpha_n}}+\alpha_n\avgD{Q_1}{Q_{\alpha_n}}\nonumber\\
    &=(1-\alpha_n)\avgD{Q_0}{Q_{\alpha_n}} + \alpha_n\avgD{Q_1}{Q_0} +\alpha_n \sum_zQ_1(z)\log \frac{Q_0(z)}{Q_{\alpha_n} (z)}\nonumber\\
    &= \alpha_n\avgD{Q_1}{Q_0} - \avgD{Q_{\alpha_n}}{Q_0}\label{eq:expression_I}
  \end{align}
  Combining~(\ref{eq:upper_bound_d}),~(\ref{eq:lower_bound_d}), and~(\ref{eq:expression_I}), we obtain the desired results.

\section{Proof of Lemma~\ref{lm:source_resolvability_div}}
\label{sec:proof-source-resolvability-div}
For any $\gamma>0$, define the set
\begin{align}
      \calS^n_\gamma &\eqdef\left\{\mathbf{x}\in\{x_0,x_1\}^n:\log \frac{1}{\Pi_{\alpha_n}^\pn(\mathbf{x})}<\gamma\right\}.
\end{align} 
For $i\in\intseq{1}{\widetilde{K}}$, we denote the expected value over all random codewords $\{\tilde{\mathbf{X}}_j\}_{j\in\intseq{1}{\widetilde{K}}\setminus\{i\}}$ by $\E[\sim i]{\cdot}$. Notice that
\begin{align}
  \E{\avgD{P_{\widetilde{\mathbf{X}}}}{\Pi_{\alpha_n}^\pn}} &= \E{\sum_{\mathbf{x}}P_{\widetilde{\mathbf{X}}}(\mathbf{x})\log\frac{P_{\widetilde{\mathbf{X}}}(\mathbf{x})}{\Pi_{\alpha_n}^\pn(\mathbf{x})}}\\
  &= \E{\sum_{\mathbf{x}}\sum_{i=1}^{\widetilde{K}}\frac{1}{{\widetilde{K}}}\mathbf{1}\left\{\mathbf{x}=\tilde{\mathbf{X}}_i\right\}\log\frac{\sum_{j=1}^{\widetilde{K}}\mathbf{1}\left\{\mathbf{x}=\tilde{\mathbf{X}}_j\right\}}{{\widetilde{K}} \Pi_{\alpha_n}^\pn(\mathbf{x})}}\\
  &= \sum_{i=1}^{\widetilde{K}}\frac{1}{{\widetilde{K}}} \sum_{\mathbf{x}}\sum_{\tilde{\mathbf{x}}_i}\Pi_{\alpha_n,\epsilon}^n(\tilde{\mathbf{x}}_i)\mathbf{1}\left\{\mathbf{x}=\tilde{\mathbf{x}}_i\right\}\E[\sim i]{\log\frac{\sum_{j=1}^{\widetilde{K}}\mathbf{1}\left\{\mathbf{x}=\tilde{\mathbf{X}}_j\right\}}{{\widetilde{K}} \Pi_{\alpha_n}^\pn(\mathbf{x})}}\\
  &\stackrel{(a)}{\leq}\sum_{i=1}^{\widetilde{K}}\frac{1}{{\widetilde{K}}} \sum_{\mathbf{x}}\sum_{\tilde{\mathbf{x}}_i}\Pi_{\alpha_n,\epsilon}^n(\tilde{\mathbf{x}}_i)\mathbf{1}\left\{\mathbf{x}=\tilde{\mathbf{x}}_i\right\}\log \E[\sim i]{\frac{\sum_{j=1}^{\widetilde{K}}\mathbf{1}\left\{\mathbf{x}=\tilde{\mathbf{X}}_j\right\}}{{\widetilde{K}} \Pi_{\alpha_n}^\pn(\mathbf{x})}}\displaybreak[0]\\
  &\stackrel{(b)}{=} \sum_{i=1}^{\widetilde{K}}\frac{1}{{\widetilde{K}}} \sum_{\mathbf{x}}\sum_{\tilde{\mathbf{x}}_i}\Pi_{\alpha_n,\epsilon}^n (\tilde{\mathbf{x}}_i)\mathbf{1}\left\{\mathbf{x}=\tilde{\mathbf{x}}_i\right\}{\log\left(\frac{\mathbf{1}\left\{\mathbf{x}=\tilde{\mathbf{x}}_i\right\}}{{\widetilde{K}} \Pi_{\alpha_n}^\pn(\mathbf{x})}+\frac{\widetilde{K}-1}{\widetilde{K}}\frac{\Pi_{\alpha_n,\epsilon}^n(\mathbf{x})}{\Pi_{\alpha_n}^\pn(\mathbf{x})}\right)}\displaybreak[0]\\
  &\leq\sum_{\mathbf{x}}\Pi_{\alpha_n,\epsilon}^n (\mathbf{{x}}){\log\left(\frac{1}{{\widetilde{K}} \Pi_{\alpha_n}^\pn(\mathbf{x})}+\frac{1}{\lambda_n}\right)}\displaybreak[0]\\
  &\leq \sum_{\mathbf{x}\notin\calS_\gamma^n}\frac{1}{\lambda_n}\Pi_{\alpha_n}^\pn (\mathbf{{x}}){\log\left(\frac{1}{{\widetilde{K}} \Pi_{\alpha_n}^\pn(\mathbf{x})}+\frac{1}{\lambda_n}\right)}+\sum_{\mathbf{x}\in\calS_\gamma^n}\Pi_{\alpha_n,\epsilon}^n (\mathbf{{x}}){\log\left(\frac{1}{{\widetilde{K}} \Pi_{\alpha_n}^\pn(\mathbf{x})}+\frac{1}{\lambda_n}\right)}\label{eq:bound_div_res_lemma_1},
\end{align}
where $(a)$ follows by Jensen's inequality and $(b)$ holds because $\E[\sim i]{\mathbf{1}\{\mathbf{x}=\smash{\tilde{\mathbf{X}}_j}\}}=\Pi_{\alpha_n,\epsilon}^n(\mathbf{x})$ for $i\neq j$.
If $\mathbf{x}\in\calS_\gamma^n$, we have $1\leq \Pi_{\alpha_n}^\pn(\mathbf{x}) e^\gamma$ and 
\begin{align}
  \log\left(\frac{1}{\widetilde{K} \Pi_{\alpha_n}^\pn(\mathbf{x})}+\frac{1}{\lambda_n}\right) \leq \log\left(\frac{e^{\gamma}}{\widetilde{K}}\frac{\Pi_{\alpha_n}^\pn(\mathbf{x})}{\Pi_{\alpha_n}^\pn(\mathbf{x})}+\frac{1}{\lambda_n}\right) = \log\left(\frac{e^{\gamma}}{\widetilde{K}}+\frac{1}{\lambda_n}\right) \label{eq:bound_div_res_lemma_2}
\end{align}
If $\mathbf{x}\notin\calS_\gamma^n$, and for $n$ large enough so that $\alpha_n<1-\alpha_n$ and $\alpha_n^n\leq \lambda_n$, we have
\begin{align}
    \log\left(\frac{1}{\widetilde{K} \Pi_{\alpha_n}^\pn(\mathbf{x})}+\frac{1}{\lambda_n}\right) \leq \log\left(\frac{1}{\alpha_n^n}+\frac{1}{\lambda_n}\right) \leq n\log\frac{2}{\alpha_n}. \label{eq:bound_div_res_lemma_3}
\end{align}
Combining~(\ref{eq:bound_div_res_lemma_1})-(\ref{eq:bound_div_res_lemma_3}), we obtain
\begin{align}
    \E{\avgD{P_{\widetilde{\mathbf{X}}}}{\Pi_{\alpha_n}^\pn}} \leq \frac{n}{\lambda_n}\log\frac{2}{\alpha_n}\P[\Pi_{\alpha_n}^\pn]{\mathbf{X}\notin \calS_\gamma^n} + \log\left(\frac{e^{\gamma}}{\widetilde{K}}+\frac{1}{\lambda_n}\right) .
\end{align}
The result follows by observing that
\begin{align}
\P[\Pi_{\alpha_n}^\pn]{\mathbf{X}\notin \calS_\gamma^n}&=  \P[\Pi_{\alpha_n}^\pn]{\log \frac{1}{\Pi_{\alpha_n}^\pn(\mathbf{X})}\geq\gamma}\\
                             &=\P[\Pi_{\alpha_n}^\pn]{\log \frac{1}{\alpha_n^{\text{supp}(\mathbf{X})}(1-\alpha_n)^{n-\text{supp}(\mathbf{X})}}\geq\gamma}\\
                             &=\P[\Pi_{\alpha_n}^\pn]{\text{supp}(\mathbf{X})\log \frac{1-\alpha_n}{\alpha_n}-n\log (1-\alpha_n)\geq\gamma}\\
                             &=\P[\Pi_{\alpha_n}^\pn]{\text{supp}(\mathbf{X})\geq\frac{\gamma+n\log (1-\alpha_n)}{\log \frac{1-\alpha_n}{\alpha_n}}}.
\end{align}

\section{Proof of Lemma~\ref{lm:reliability-lpd}}
\label{sec:proof-lemma-reliability-lpd}
The result of the lemma could be viewed as a specific application of the $\kappa\beta$ bound~\cite{Polyanskiy2010}. However, for clarity and completeness, we provide here a proof from first principles. From the definition of the encoder/decoder and a union bound, there are three error events to consider.
\begin{itemize}
\item The codeword $\mathbf{x}_{ij}$ is transmitted but $(\mathbf{x}_{ij},\mathbf{y})\notin\calA_\gamma^n$.
\item The codeword $\mathbf{x}_{ij}$ is transmitted but there exists $\mathbf{x}_{kj}$ with $k\neq i$ such that $(\mathbf{x}_{kj},\mathbf{y})\in\calA_\gamma^n$.
\item No communication happens but the decoder finds a codeword $\mathbf{x}_{ij}$ such that $(\mathbf{x}_{ij},\mathbf{y})\in\calA_\gamma^n$.
\end{itemize}
Hence, we obtain
  \begin{align}
    \E{P_{\text{err}}} &\leq\E{\sum_{\mathbf{y}}\sum_{i=1}^M\sum_{j=1}^{K}\frac{1}{MK}W_{Y|X}^\pn(\mathbf{y}|\mathbf{X}_{ij})\mathbf{1}\left\{(\mathbf{X}_{ij},\mathbf{y})\notin\calA_{\gamma}^n\text{ or }\exists k\neq i \text{ s.t. }(\mathbf{X}_{kj},\mathbf{y})\in\calA_{\gamma}^n\right\}}\nonumber\\
    &\phantom{=========}+\E{\sum_{\mathbf{y}}P_{0}^\pn(\mathbf{y})\mathbf{1}\left\{\exists i \text{ s.t. }(\mathbf{X}_{ij},\mathbf{y})\in\calA_{\gamma}^n\right\}}\nonumber\displaybreak[0]\\
            &= \E{\sum_{\mathbf{y}} W_{Y|X}^\pn(\mathbf{y}|\mathbf{X}_{11})\mathbf{1}\left\{(\mathbf{X}_{11},\mathbf{y})\notin\calA_{\gamma}^n\text{ or }\exists k\neq 1 \text{ s.t. }(\mathbf{X}_{k1},\mathbf{y})\in\calA_{\gamma}^n\right\}}\nonumber\\
    &\phantom{=========}+\E{\sum_{\mathbf{y}}P_{0}^\pn(\mathbf{y})\mathbf{1}\left\{\exists i \text{ s.t. }(\mathbf{X}_{ij},\mathbf{y})\in\calA_{\gamma}^n\right\}}\displaybreak[0]\nonumber\\
            & \leq \E{\sum_{\mathbf{y}} W_{Y|X}^\pn(\mathbf{y}|\mathbf{X}_{11})\mathbf{1}\left\{(\mathbf{X}_{11},\mathbf{y})\notin\calA_{\gamma}^n\right\}}+\sum_{k\neq 1}\E{\sum_{\mathbf{y}}W_{Y|X}^\pn(\mathbf{y}|\mathbf{X}_{11})\mathbf{1}\left\{(\mathbf{X}_{k1},\mathbf{y})\in\calA_{\gamma}^n\right\}}\nonumber\\
    &\phantom{=========}+\sum_i \E{\sum_{\mathbf{y}}P_{0}^\pn(\mathbf{y})\mathbf{1}\left\{(\mathbf{X}_{ij},\mathbf{y})\in\calA_{\gamma}^n\right\}}\label{eq:bound_avg_pe}
  \end{align}
  Note that the first term on the right-hand side of~(\ref{eq:bound_avg_pe}) is
  \begin{align}
        \E{\sum_{\mathbf{y}} W_{Y|X}^\pn(\mathbf{y}|\mathbf{X}_{11})\mathbf{1}\left\{(\mathbf{X_{11}},\mathbf{y})\notin\calA_{\gamma}^n\right\}} &= \P[W_{Y|X}^\pn\Pi_{\alpha_n}^\pn]{\log \frac{W_{Y|X}^\pn(\mathbf{Y}|\mathbf{X})}{P_0^\pn(\mathbf{Y})}\leq\gamma}.
  \end{align}
  We now analyze the second term on the right-hand side of~(\ref{eq:bound_avg_pe}). For any $k\neq 1$,
  \begin{align}
    \E{\sum_{\mathbf{y}} W_{Y|X}^\pn(\mathbf{y}|\mathbf{X}_{11})\mathbf{1}\left\{(\mathbf{X}_{k1},\mathbf{y})\in\calA_{\gamma}^n\right\}} &= \sum_{\mathbf{x}}\sum_{\mathbf{y}} P_{\alpha_n}^\pn(\mathbf{y}) \Pi_{\alpha_n}^\pn(\mathbf{x}) \mathbf{1}\left\{(\mathbf{x},\mathbf{y})\in\calA_{\gamma}^n\right\}\\\displaybreak[0]
 &= \sum_{\mathbf{x}}\sum_{\mathbf{y}} P_0^\pn(\mathbf{y}) \Pi_{\alpha_n}^\pn(\mathbf{x}) \frac{P_{\alpha_n}^\pn(\mathbf{y})}{P_0^\pn(\mathbf{y})}\mathbf{1}\left\{(\mathbf{x},\mathbf{y})\in\calA_{\gamma}^n\right\}\\
 &\stackrel{(a)}{\leq} \sum_{\mathbf{x}}\sum_{\mathbf{y}} W_{Y|X}^\pn(\mathbf{y}|\mathbf{x}) e^{-\gamma}\Pi_{\alpha_n}^\pn(\mathbf{x}) \frac{P_{\alpha_n}^\pn(\mathbf{y})}{P_0^\pn(\mathbf{y})}\mathbf{1}\left\{(\mathbf{x},\mathbf{y})\in\calA_{\gamma}^n\right\}\\
 &\leq e^{-\gamma}\E[P_{\alpha_n}^\pn]{\frac{P_{\alpha_n}^\pn(\mathbf{Y})}{P_0^\pn(\mathbf{Y})}},
  \end{align}
where $(a)$ follows from $P_0^\pn(\mathbf{y})\leq W_{Y|X}^\pn(\mathbf{y}|\mathbf{x}) e^{-\gamma}$ for $(\mathbf{x},\mathbf{y})\in\calA_{\gamma}^n$. Since $P_{\alpha_n}^\pn$ and $P_0^\pn$ are product distributions, we have
  \begin{align}
    \E[P_{\alpha_n}^\pn]{\frac{P_{\alpha_n}^\pn(\mathbf{Y})}{P_0^\pn(\mathbf{Y})}} =\left(\E[P_{\alpha_n}]{\frac{P_{\alpha_n}(Y)}{P_0(Y)}}\right)^n
  \end{align}
  Next, note that
  \begin{align}
    \E[P_{\alpha_n}]{\frac{P_{\alpha_n}(Y)}{P_0(Y)}}&=\E[P_{\alpha_n}]{1-\alpha_n+\alpha_n\frac{P_1(Y)}{P_0(Y)}}\\
                                  &=1-\alpha_n + \alpha_n\left(\sum_{y}\left((1-\alpha_n)P_0(y)+\alpha_nP_1(y)\right) \frac{P_1(y)}{P_0(y)}\right)\\
                                  &=1-\alpha_n + \alpha_n\left(1-\alpha_n+\alpha_n\sum_{y}\frac{P_1(y)^2}{P_0(y)}\right)\\
                                  &=1+\alpha_n^2(\zeta-1),\text{ with $\zeta\eqdef\sum_{y}\frac{P_1(y)^2}{P_0(y)}$.}
  \end{align}
  Consequently,
  \begin{align}
    \E[P_{\alpha_n}^\pn]{\frac{P_{\alpha_n}^\pn(\mathbf{Y})}{P_0^\pn(\mathbf{Y})}} = \left(1+\alpha_n^2(\zeta-1)\right)^n =\exp\left(n\log \left(1+\alpha_n^2(\zeta-1)\right)\right) \leq \exp\left(n\alpha_n^2(\zeta-1)\right)&= \exp\left(\omega_n^2(\zeta-1)\right).
  \end{align}
  Hence, we obtain
  \begin{align}
    \E{\sum_{\mathbf{y}} W_{Y|X}^\pn(\mathbf{y}|\mathbf{X}_{11})\mathbf{1}\left\{(\mathbf{X}_{k1},\mathbf{y})\in\calA_{\gamma}^n\right\}} \leq e^{-\gamma}\exp\left(\omega_n^2(\zeta-1)\right). \label{eq:bound_avg_pe_1}
  \end{align}
  Finally, the third term on the right-hand side of~(\ref{eq:bound_avg_pe}) may be similarly bounded for any $i$ by
  \begin{align}
    \E{\sum_{\mathbf{y}}P_{0}^\pn(\mathbf{y})\mathbf{1}\left\{(\mathbf{X}_{i1},\mathbf{y})\in\calA_{\gamma}^n\right\}}&=\sum_{\mathbf{x}}\sum_{\mathbf{y}}P_0^\pn(\mathbf{y})\Pi_{\alpha_n}^\pn(\mathbf{x}) \mathbf{1}\left\{(\mathbf{x},\mathbf{y})\in\calA_{\gamma}^n\right\}\\
    &\leq \sum_{\mathbf{x}}\sum_{\mathbf{y}}W_{Y|X}^\pn(\mathbf{y}|\mathbf{x})e^{-\gamma}\Pi_{\alpha_n}^\pn(\mathbf{x}) \mathbf{1}\left\{(\mathbf{x},\mathbf{y})\in\calA_{\gamma}^n\right\}\\
    &\leq e^{-\gamma}.
  \end{align}

\section{Proof of Lemma~\ref{lm:lemma-resolvability-lpd-div}}
\label{sec:proof-lemma-resovlability-lpd-div}
Define the set
\begin{align}
    \calB^n_\tau&\eqdef\left\{(\mathbf{x},\mathbf{z})\in\calX^n\times\calZ^n:\log \frac{W_{Z|X}^\pn(\mathbf{z}|\mathbf{x})}{Q_0^\pn(\mathbf{z})}<\tau\right\}
\end{align} 
For $(i,j)\in\intseq{1}{M}\times\intseq{1}{K}$, we denote the expected value over all random codewords $\{\tilde{\mathbf{X}}_{k\ell}\}_{(k,\ell)\in\intseq{1}{M}\times\intseq{1}{K}\setminus\{(i,j)\}}$ by $\E[\sim ij]{\cdot}$. Notice that
\begin{align}
  \E{\avgD{\hQ^n}{Q_{\alpha_n}^\pn}} &= \E{\sum_{\mathbf{z}} \hQ^n(\mathbf{z})\log\frac{\hQ^n(\mathbf{z})}{Q_{\alpha_n}^\pn(\mathbf{z})}}\nonumber\\
  &=\E{\sum_{\mathbf{z}} \frac{1}{MK}\sum_{i=1}^{M}\sum_{j=1}^{K}  W_{Z|X}^\pn(\mathbf{z}|\mathbf{X}_{ij})\log\frac{\sum_{k=1}^{M}\sum_{\ell=1}^{K} W_{Z|X}^\pn(\mathbf{z}|\mathbf{X}_{k\ell})}{MKQ_{\alpha_n}^\pn(\mathbf{z})}}\nonumber\\
  &=\frac{1}{{MK}}\sum_{i=1}^{M}\sum_{j=1}^K\sum_{\mathbf{z}} \sum_{\mathbf{x}_{ij}}W_{Z|X}^\pn(\mathbf{z}|\mathbf{x}_{ij})\Pi_{\alpha_n}^\pn(\mathbf{x}_{ij})\E[\sim ij]{\log\frac{\sum_{k=1}^{M}\sum_{\ell=1}^{K} W_{Z|X}^\pn(\mathbf{z}|\mathbf{X}_{k\ell})}{MKQ_{\alpha_n}^\pn(\mathbf{z})}}\nonumber\\
  &\stackrel{(a)}{\leq} \frac{1}{{MK}}\sum_{i=1}^{M}\sum_{j=1}^K\sum_{\mathbf{z}} \sum_{\mathbf{x}_{ij}}W_{Z|X}^\pn(\mathbf{z}|\mathbf{x}_{ij})\Pi_{\alpha_n}^\pn(\mathbf{x}_{ij})\log \E[\sim ij]{\frac{\sum_{k=1}^{M}\sum_{\ell=1}^{K} W_{Z|X}^\pn(\mathbf{z}|\mathbf{X}_{k\ell})}{MKQ_{\alpha_n}^\pn(\mathbf{z})}}\nonumber\\
                                     &\stackrel{(b)}{=} \frac{1}{{MK}}\sum_{i=1}^{M}\sum_{j=1}^K\sum_{\mathbf{z}} \sum_{\mathbf{x}_{ij}}W_{Z|X}^\pn(\mathbf{z}|\mathbf{x}_{ij})\Pi_{\alpha_n}^\pn(\mathbf{x}_{ij})\log \left(\frac{W_{Z|X}^\pn(\mathbf{z}|\mathbf{x}_{ij})}{{MK}Q_{\alpha_n}^\pn(\mathbf{z})}+\frac{{MK}-1}{{MK}}\right)\nonumber\\
&=  \sum_{\mathbf{z}} \sum_{\mathbf{x}}W_{Z|X}^\pn(\mathbf{z}|\mathbf{x})\Pi_{\alpha_n}^\pn(\mathbf{x})\log \left(\frac{W_{Z|X}^\pn(\mathbf{z}|\mathbf{x})}{{MK}Q_{\alpha_n}^\pn(\mathbf{z})}+\frac{{MK}-1}{{MK}}\right),\label{eq:bound_div_1}
\end{align}
where $(a)$ follows by Jensen's inequality and $(b)$ follow because $\E[\sim ij]{W_{Z|X}^\pn(\mathbf{z}|\mathbf{X}_{k\ell})}=Q_{\alpha_n}^\pn(\mathbf{z})$ for $(k,\ell)\neq(i,j)$.
If $(\mathbf{x},\mathbf{z})\in\calB_\tau^n$, we have
\begin{align}
  \log \left(\frac{W_{Z|X}^\pn(\mathbf{z}|\mathbf{x})}{{MK}Q_{\alpha_n}^\pn(\mathbf{z})}+\frac{{MK}-1}{{MK}}\right) \leq   \log \left(\frac{e^{\tau}Q_0^\pn(\mathbf{z})}{{MK}Q_{\alpha_n}^\pn(\mathbf{z})}+1\right)\leq \frac{e^{\tau}}{{MK}}\frac{Q_0^\pn(\mathbf{z})}{Q_{\alpha_n}^\pn(\mathbf{z})}.\label{eq:bound_div_2}
\end{align}
If $(\mathbf{x},\mathbf{z})\notin\calB_\tau^n$, we have
\begin{align}
  \log \left(\frac{W_{Z|X}^\pn(\mathbf{z}|\mathbf{x})}{{MK}Q_{\alpha_n}^\pn(\mathbf{z})}+\frac{{MK}-1}{{MK}}\right) \leq   \log \left(\frac{1}{(1-\alpha_n)^n\mu_0^n}+1\right)\leq n\log\frac{2}{(1-\alpha_n)\mu_0}.\label{eq:bound_div_3}
\end{align}
Combining~(\ref{eq:bound_div_1})-(\ref{eq:bound_div_3}), we obtain
\begin{align*}
    \E{\avgD{\hQ^n}{Q_{\alpha_n}^\pn}}&\leq n\log\frac{2}{(1-\alpha_n)\mu_0} \sum_{\mathbf{z}} \sum_{\mathbf{x}}W_{Z|X}^\pn(\mathbf{z}|\mathbf{x})\Pi_{\alpha_n}^\pn(\mathbf{x})\mathbf{1}\{(\mathbf{x},\mathbf{z})\notin\calB_\tau^n\}\\
  &\phantom{===============}+  \sum_{\mathbf{z}} \sum_{\mathbf{x}}W_{Z|X}^\pn (\mathbf{z}|\mathbf{x})\Pi_{\alpha_n}^\pn(\mathbf{x}) \frac{e^{\tau}}{{MK}}\frac{Q_0^\pn(\mathbf{z})}{Q_{\alpha_n}^\pn(\mathbf{z})}\mathbf{1}\{(\mathbf{x},\mathbf{z})\in\calB_\tau^n\}\\
                                              &\leq n\log\frac{2}{(1-\alpha_n)\mu_0} \P[W_{Z|X}^\pn\Pi_{\alpha_n}^\pn]{(\mathbf{X},\mathbf{Z})\notin\calB_\tau^n} + \frac{e^{\tau}}{MK}.
\end{align*}
For $n$ large enough so that $1-\alpha_n\geq 1/2$, we obtain the desired result.

\section{Proof of Lemma~\ref{lm:lemma-resolvability-lpd-var}}
\label{sec:proof-lemma-resovability-lpd-var}

We define
\begin{align}
  \hQ^{(1)}(\mathbf{z})&\eqdef \sum_{i=1}^M\sum_{j=1}^KW_{Z|X}^n(\mathbf{z}|\mathbf{x}_{ij})\frac{1}{MK}\mathbf{1}\{(\mathbf{x}_{ij},\mathbf{z})\in\calB^n_\tau\}\\
  \hQ^{(2)}(\mathbf{z})&\eqdef \sum_{i=1}^M\sum_{j=1}^KW_{Z|X}^n(\mathbf{z}|\mathbf{x}_{ij})\frac{1}{MK}\mathbf{1}\{(\mathbf{x}_{ij},\mathbf{z})\notin\calB^n_\tau\}
\end{align}
so that $\hQ^n=\hQ^{(1)}+\hQ^{(2)}$. Also note that $  \E{\hQ(\mathbf{z})} = Q_{\alpha_n}^\pn(\mathbf{z})$. Hence,
\begin{align}
  \E{\Vert \hQ^n-Q_{\alpha_n}^\pn\Vert} &\leq \frac{1}{2}\sum_{\mathbf{z}}\E{\abs{  \hQ^{(1)}(\mathbf{z})-\E{  \hQ^{(1)}(\mathbf{z})}}}+\frac{1}{2}\sum_{\mathbf{z}}\E{\abs{  \hQ^{(2)}(\mathbf{z})-\E{  \hQ^{(2)}(\mathbf{z})}}}.\label{eq:bound_var_dist_bis}
\end{align}
The first term on the right-hand side of~(\ref{eq:bound_var_dist_bis}) is bounded as follows.
\begin{align}
  \frac{1}{2}\sum_{\mathbf{z}}\E{\abs{  \hQ^{(1)}(\mathbf{z})-\E{  \hQ^{(1)}(\mathbf{z})}}} & \leq \frac{1}{2}\sum_{\mathbf{z}}\sqrt{\Var{\hQ^{(1)}(\mathbf{z})}}
\end{align}
with
\begin{align}
  \Var{\hQ^{(1)}(\mathbf{z})} &= \sum_{i=1}^M \sum_{j=1}^K \frac{1}{M^2K^2}\Var{W_{Z|X}^\pn(\mathbf{z}|\mathbf{X}_{ij})\mathbf{1}\{(\mathbf{X}_{ij},\mathbf{z})\in\calB^n_\tau\}}\\
  &=\frac{1}{MK}\Var{W_{Z|X}^\pn(\mathbf{z}|\mathbf{X}_{11})\mathbf{1}\{(\mathbf{X}_{11},\mathbf{z})\in\calB^n_\tau\}}\\
  &\leq \frac{1}{MK}\E[\Pi_{\alpha_n}^\pn]{W_{Z|X}^\pn(\mathbf{z}|\mathbf{X})^2\mathbf{1}\{(\mathbf{X},\mathbf{z})\in\calB^n_\tau\}}\\
  &= \frac{1}{MK}\sum_{\mathbf{x}}W_{Z|X}^\pn(\mathbf{z}|\mathbf{x})^2\Pi_{\alpha_n}^\pn(\mathbf{x}) \mathbf{1}\{(\mathbf{x},\mathbf{z})\in\calB^n_\tau\}\\
  &\stackrel{(a)}{\leq} \frac{1}{MK}\sum_{\mathbf{x}}W_{Z|X}^\pn(\mathbf{z}|\mathbf{x})Q_0^\pn(\mathbf{z})e^{\tau}\Pi_{\alpha_n}^\pn(\mathbf{x}) \mathbf{1}\{(\mathbf{x},\mathbf{z})\in\calB^n_\tau\}\\
  &\leq\frac{1}{MK} Q_0^\pn(\mathbf{z})^2 e^{\tau} \frac{Q_{\alpha_n}^\pn(\mathbf{z})}{Q_0^\pn(\mathbf{z})},
\end{align}
where $(a)$ follows because $W_{Z|X}^\pn(\mathbf{z}|\mathbf{x})\leq Q_0^\pn(\mathbf{z})e^{\tau}$ for $(\mathbf{x},\mathbf{z})\in\calB^n_\tau$. Hence,
\begin{align}
    \frac{1}{2}\sum_{\mathbf{z}}\E{\abs{  \hQ^{(1)}(\mathbf{z})-\E{  \hQ^{(1)}(\mathbf{z})}}} & \leq \frac{1}{2}\sum_{\mathbf{z}}\sqrt{\frac{1}{MK} Q_0^\pn(\mathbf{z})^2 e^{\tau} \frac{Q_{\alpha_n}^\pn(\mathbf{z})}{Q_0^\pn(\mathbf{z})}}
  = \frac{1}{2}\sqrt{\frac{e^\tau}{MK}}\sum_{\mathbf{z}}Q_0^\pn(\mathbf{z})\sqrt{\frac{Q_{\alpha_n}^\pn(\mathbf{z})}{Q_0^\pn(\mathbf{z})}}.
\end{align}
By Jensen's inequality and the concavity of $x\mapsto \sqrt{x}$, we have
\begin{align}
  \sum_{\mathbf{z}}Q_0^\pn(\mathbf{z})\sqrt{\frac{Q_{\alpha_n}^\pn(\mathbf{z})}{Q_0^\pn(\mathbf{z})}}\leq \sqrt{  \sum_{\mathbf{z}} Q_{\alpha_n}^\pn(\mathbf{z})} =1,
\end{align}
so that
\begin{align}
  \frac{1}{2}\sum_{\mathbf{z}}\E{\abs{  \hQ^{(1)}(\mathbf{z})-\E{  \hQ^{(1)}(\mathbf{z})}}} \leq \frac{1}{2}\sqrt{\frac{e^\tau}{M}}\label{eq:bound_var_dist_1}.
\end{align}
The second term on the right-hand side of~(\ref{eq:bound_var_dist_bis}).
\begin{align}
  \frac{1}{2}\sum_{\mathbf{z}}\E{\abs{  \hQ^{(2)}(\mathbf{z})-\E{  \hQ^{(2)}(\mathbf{z})}}}&\leq \sum_{\mathbf{z}}\E{\hQ^{(2)}(\mathbf{z})}\\
  &=\sum_{\mathbf{z}}\sum_{i=1}^M \sum_{j=1}^K \sum_{\mathbf{x}_{ij}} W_{Z|X}^\pn(\mathbf{z}|\mathbf{x}_{ij})\Pi_{\alpha_n}^\pn(\mathbf{x}_{ij})\frac{1}{MK}\mathbf{1}\{(\mathbf{x}_{ij},\mathbf{z})\notin\calB^n_\tau\}\\
  &=\sum_{\mathbf{z}}\sum_{\mathbf{x}}W_{Z|X}^\pn(\mathbf{z}|\mathbf{x})\Pi_{\alpha_n}^\pn(\mathbf{x})\mathbf{1}\{(\mathbf{x},\mathbf{z})\notin\calB^n_\tau\}\\
  &=\P[W_{Z|X}^\pn\Pi_{\alpha_n}^\pn]{\log \frac{W_{Z|X}^\pn(\mathbf{Z}|\mathbf{X})}{Q_0^\pn(\mathbf{Z})}>\tau}.
\end{align}

\section{Special cases of channels}
\label{sec:spec-cases-chann}
In this Appendix, we discuss some special cases of channels that have been excluded by the assumptions $P_1\ll P_0$, $Q_1\ll Q_0$, and $Q_1\neq Q_0$, made in Section~\ref{sec:covert-comm-over}.

\subsection{$Q_1$ is not absolutely continuous \ac{wrt} $Q_0$ or $Q_1=Q_0$}
\label{sec:ward-chann-which}

If $Q_1$ is not absolutely continuous \ac{wrt} $Q_0$ then $\avgD{Q_1}{Q_0}=\infty$. Hence, for any $n\in\mathbb{N}^*$ and any sequence $\mathbf{x}\in\{x_0,x_1\}^n$ distinct from the all-$x_0$ sequence, we have
\begin{align}
  \avgD{W_{Z^n|X^n=\mathbf{x}}}{Q_0^{\pn}}=\sum_{i=1}^n\avgD{W_{Z|X=x_i}}{Q_0} =\infty.
\end{align}
Consequently, it is impossible to transmit covert bits.

If $Q_1=Q_0$, for any $n\in\mathbb{N}^*$ and any transmitted sequence $\mathbf{x}\in\{x_0,x_1\}^n$, we have $\avgD{W_{Z^n|X^n=\mathbf{x}}}{Q_0^{\pn}}=0$, i.e., the warden's observations are independent of the transmitted signals and always have distribution $Q_0^\pn$. One may therefore use a standard error-control code for reliability over the main channel and transmit at non-vanishing rates approaching the capacity of the main channel. The corresponding scaling of $\log M$ is $\Theta(n)$.

\subsection{$P_1$ is not absolutely continuous \ac{wrt} $P_0$}
\label{sec:main-channels-which}

If $P_1$ is not absolutely continuous \ac{wrt} $P_0$, denoted $P_1\nll P_0$, define
\begin{align}
  \calS\eqdef \{y\in\calY: P_1(y)> 0\text{ and }P_0(y)=0\}\quad\text{and}\quad \kappa\eqdef\sum_{y\in\calS}P_1(y).\label{eq:def_prob_no_error}
\end{align}
In other words, $\kappa$ is the probability that the symbol $x_1$ is identified without ambiguity at the channel output. We then have the following.

\begin{theorem}
    Consider a discrete memoryless covert communication channel with $P_1\nll P_0$, $Q_1\ll Q_0$, and $Q_1\neq Q_0$. Let $\kappa$ be defined as per~(\ref{eq:def_prob_no_error}) and let $\alpha_n\eqdef\frac{\omega_n}{\sqrt{n}}$ with $\omega_n\in o(1)\cap\omega(\frac{1}{\sqrt{n}})$ as $n\rightarrow \infty$. For any $\xi\in]0;1[$, there exist $\xi_1, \xi_2>0$ depending on $\xi$, $W_{Y|X}$, $W_{Z|X}$, and covert communication schemes such that, for all $n$ large enough,
  \begin{align*}
    \log M &= (1-\xi)\kappa\left(\frac{1}{2}+\frac{\log \omega_n^{-1}}{\log n}\right)\omega_n\sqrt{n}\log n,\qquad\log K=0,
  \end{align*}
  and
\begin{align*}
  P_{\text{err}}\leq e^{-\xi_1\omega_n\sqrt{n}},\qquad    \abs{\avgD{\hQ^n}{Q_{0}^\pn}-\avgD{Q_{\alpha_n}^\pn}{Q_{0}^\pn}} \leq e^{-\xi_2\omega_n\sqrt{n}}.
  \end{align*}
\end{theorem}

\begin{IEEEproof}
  The result follows with a modification of the proof of Theorem~\ref{th:key-assist-2} to exploit the property $P_1\nll P_0$. Let $\delta>0$, $M\in\mathbb{N}^*$, and $\alpha_n
\eqdef\frac{\omega_n}{\sqrt{n}}$. Generate $M$ codewords $\mathbf{x}_i$ with $i\in\intseq{1}{M}$ independently according to the product distribution $\Pi_{\alpha_n}^\pn$. Upon receiving $\mathbf{y}$, the decoder looks for symbols that belong to $\calS$. Let $\calP(\mathbf{y})$ denote the positions of these symbols. Then,
  \begin{itemize}
  \item if $\card{\calP(\mathbf{y})}<(1-\delta) n \kappa \alpha_n$ declare that $\hT=0$;
  \item else, if there exists a unique $i\in\intseq{1}{M}$ such that codeword $\mathbf{x}_i$ has $x_1$-symbols for all positions in $\calP(\mathbf{y})$, declare $\hT=1$ and output message $\hW=i$;
  \item otherwise, declare an error.
  \end{itemize}

  \paragraph{Channel reliability analysis} By construction, the decoder makes an error if any of the following events occur.
  \begin{itemize}
  \item The codeword $\mathbf{x}_i$ is transmitted but there are fewer than $(1-\delta) n \kappa \alpha_n$ $x_1$-symbols in $\calP(\mathbf{y})$.
  \item The codeword $\mathbf{x}_i$ is transmitted but there are multiple codewords with $x_1$-symbols for all positions in $\calP(\mathbf{y})$.
  \item No communication takes place but there are more than $(1-\delta) n \kappa \alpha_n$ $x_1$-symbols in $\calP(\mathbf{y})$.
  \end{itemize}
By definition of $\calS$, note that $\card{\calP(\mathbf{y})}=0$ if no communication takes place. Consequently, the probability of error averaged over the random codebook generation satisfies
  \begin{align}
    \E{P_{\text{err}}} &\leq \E{\sum_{\mathbf{y}}\sum_{i=1}^M\frac{1}{M}W_{Y|X}^\pn(\mathbf{y}|\mathbf{X}_i) \mathbf{1}\{\card{\calP(\mathbf{y})}<(1-\delta) n \kappa \alpha_n\text{ or }\exists j\neq i \text{ such that } \forall k\in\calS\;X_{j,k}=X_{i,k}\}}\\
            & \leq \E{\sum_{\mathbf{y}}W_{Y|X}^\pn(\mathbf{y}|\mathbf{X}_1) \mathbf{1}\{\card{\calP(\mathbf{y})}<(1-\delta) n \kappa \alpha_n\}}\nonumber\\
            &\phantom{==========}+ \sum_{j\neq 1}\E{\sum_{\mathbf{y}}W_{Y|X}^\pn(\mathbf{y}|\mathbf{X}_1)\mathbf{1}\{\card{\calP(\mathbf{y})}\geq(1-\delta) n \kappa \alpha_n\text{ and }\forall k\in\calS\;X_{j,k}=X_{1,k}\}}\displaybreak[0]\\
            &= \P[P_{\alpha_n}^\pn]{\card{\calP(\mathbf{Y})}<(1-\delta) n \kappa \alpha_n}\nonumber\\
            &\phantom{======}+ \sum_{j\neq 1}\sum_{\mathbf{x}}\sum_{\mathbf{x'}}\sum_{\mathbf{y}}\Pi_{\alpha_n}^\pn(\mathbf{x}) \Pi_{\alpha_n}^\pn(\mathbf{x'}) W_{Y|X}^\pn(\mathbf{y}|\mathbf{x})\mathbf{1}\{\card{\calP(\mathbf{y})}\geq(1-\delta) n \kappa \alpha_n\text{ and }\forall k\in\calS\;x'_{k}=x_{k}\}\\
            &\leq \P[P_{\alpha_n}^\pn]{\card{\calP(\mathbf{Y})}<(1-\delta) n \kappa \alpha_n}+ \sum_{j\neq 1}\sum_{\mathbf{x}}\sum_{\mathbf{y}}\Pi_{\alpha_n}^\pn(\mathbf{x}) W_{Y|X}^\pn(\mathbf{y}|\mathbf{x})\mathbf{1}\{\card{\calP(\mathbf{y})}\geq(1-\delta) n \kappa \alpha_n\}\alpha_n^{\card{\calP(\mathbf{y})}}\\
            &\leq \P[P_{\alpha_n}^\pn]{\card{\calP(\mathbf{Y})}<(1-\delta) n \kappa \alpha_n} + M \alpha_n^{(1-\delta) n \kappa \alpha_n}.
  \end{align}
  Since
  $\card{\calP(\mathbf{Y})}=\sum_{i=1}^n\mathbf{1}\{Y_i\in\calS\}$ and $\E[P_{\alpha_n}^\pn]{\card{\calP(\mathbf{Y})}}=\alpha_nn\kappa=\omega_n\sqrt{n}\kappa$, a
  Chernoff bound guarantees that
  \begin{align}
    \P[P_{\alpha_n}^\pn]{\card{\calP(\mathbf{Y})}<(1-\delta) \alpha_n n \kappa }  \leq e^{-\frac{\delta^2}{2}\omega_n\sqrt{n}\kappa}.
  \end{align}
  Hence, for any $\mu\in]0;1[$, choosing
  \begin{align}
    \log M = (1-\mu)(1-\delta) \kappa\left(\frac{1}{2}+\frac{\log\omega_n^{-1}}{\log n}\right) \omega_n\sqrt{n}\log n\label{eq:choice_M_n_logn}
  \end{align}
  ensures that
  \begin{align}
    \E{P_{\text{err}}} \leq e^{-\rho_1\omega_n\sqrt{n}}\text{ for some appropriate choice of $\rho_1>0$}.
  \end{align}

  \paragraph{Channel resolvability analysis}
  Lemma~\ref{lm:lemma-resolvability-lpd-div} still applies and one may pursue the same analysis as in the proof of Theorem~\ref{th:key-assist-2}. In fact, the choice of $\log M$ in~(\ref{eq:choice_M_n_logn}) is overwhelmingly larger than the minimum required to ensure
  \begin{align}
    \E{\avgD{\hQ^n}{Q_{\alpha_n}^\pn}}\leq e^{-\rho_2\omega_n\sqrt{n}}\text{ for some appropriate choice of $\rho_2>0$}.
  \end{align}
  Note that this may be achieved without using any secret key. The final steps of the proof are identical to those in the proof of Theorem~\ref{th:key-assist-2}.
\end{IEEEproof}

We may also identify the corresponding asymptotic scaling constant of $\log M$.

\begin{corollary}
  \label{cor:scaling_p1_nac_p0}
    Consider a discrete memoryless covert communication channel with $P_1\nll P_0$, $Q_1\ll Q_0$, and $Q_1\neq Q_0$. Let $\kappa$ be defined as per~(\ref{eq:def_prob_no_error}) and $\omega_n\in o(1)\cap\omega(\frac{1}{\sqrt{n}})$ as $n\rightarrow\infty$. Then, for any $\xi\in]0;1[$, there exist keyless covert communication schemes such that
  \begin{align*}
    \lim_{n\rightarrow\infty}&\avgD{\hQ^n}{Q_0^\pn}=0, \quad \lim_{n\rightarrow\infty}P_{\text{err}}=0,\\
    \lim_{n\rightarrow\infty}&\frac{\log M}{\sqrt{n\avgD{{\hQ}^n}{Q_0^\pn}}\log n} =  (1-\xi)\kappa\sqrt{\frac{2}{\chid{Q_1}{Q_0}}} \left(\frac{1}{2}+\lim_{n\rightarrow\infty}\frac{\log\omega_n^{-1}}{\log n}\right).
  \end{align*}
\end{corollary}
\begin{IEEEproof}
  Follows from steps identical to the proof of Corollary~\ref{cor:scaling-key-assisted}.
\end{IEEEproof}
Notice that the optimal scaling constant depends on the exact choice of $\omega_n\in o(1)\cap\omega(\frac{1}{\sqrt{n}})$. This differs from the situation of Corollary~\ref{cor:scaling-conditions}, in which the scaling constant remains the same for \emph{all} choices of $\omega_n\in o(1)\cap\omega(\frac{1}{\sqrt{n}})$.\smallskip

This coding scheme turns out to be optimal, as one may establish the following converse result.
\begin{theorem}
  \label{th:converse_p1_non_ac_p0}
    Consider a discrete memoryless covert communication channel with $P_1\nll P_0$, $Q_1\ll Q_0$, and $Q_1\neq Q_0$. Consider a sequence of covert communication schemes with increasing blocklength $n$ characterized by $\epsilon_n\eqdef P_{\text{err}}$ and $\delta_n\eqdef \avgD{\hQ^n}{Q_0^\pn}$.  If $\lim_{n\rightarrow\infty}M=\infty$ and $\lim_{n\rightarrow\infty}\epsilon_n=\lim_{n\rightarrow\infty}\delta_n=0$, there exists $\varpi_n\in o(1)\cap\omega(\frac{1}{\sqrt{n}\log n})$ as $n\rightarrow\infty$ such that 
    \begin{align*}
      \lim_{n\rightarrow\infty}\frac{\log M}{\sqrt{n\avgD{{\hQ}^n}{Q_0^\pn}}\log n} \leq \kappa\sqrt{\frac{2}{\chid{Q_1}{Q_0}}}\left(\frac{1}{2}+\lim_{n\rightarrow\infty}\frac{\log\varpi_n^{-1}}{\log n}\right)
    \end{align*}
\end{theorem} 

\begin{IEEEproof} The converse proof technique of Theorem~\ref{th:converse} applies but we cannot rely on Lemma~\ref{lm:preliminary} to bound $\avgI{\smash{\tilde{X};\tilde{Y}}}$ with $\tilde{X}\sim\Pi_{\mu_n}$ since $P_1\nll P_0$. We use instead the following bound.
\begin{align}
                            \avgI{\smash{\tilde{X};\tilde{Y}}} &= (1-\mu_n)\avgD{P_0}{P_{\mu_n}} +\mu_n \avgD{P_1}{P_{\mu_n}}\\
                                                               &=(1-\mu_n)\sum_{y\in\calY\setminus\calS} P_{0}(y) \log\frac{P_{0}(y)}{P_{\mu_n}(y)} + \mu_n\sum_{y\in\calY\setminus\calS}P_{1}(y) \log\frac{P_{1}(y)}{P_{\mu_n}(y)} - \mu_n \kappa \log \mu_n\\
                             &\leq \log\frac{1}{1-\mu_n} + \mu_n \sum_{y\in\calY\setminus\calS} P_{1}(y)\log\frac{P_1(y)}{P_0(y)}+\mu_n \kappa \log \mu_n^{-1},\label{eq:bound_info_n_logn}
\end{align}
where the last inequality follows because $P_{\mu_n}(y)\geq (1-\mu_n)P_0(y)$. Since $\lim_{n\rightarrow\infty}{\sqrt{n}\mu_n}=0$ for the same reason as in the proof of Theorem~\ref{th:converse}, we may write $\mu_n=\frac{\varpi_n}{\sqrt{n}}$ with $\varpi_n=o(1)$. If $\lim_{n\rightarrow\infty}\log M=\infty$ then~(\ref{eq:converse_bound_logM_reliability}) and~(\ref{eq:bound_info_n_logn}) impose that
\begin{align}
\lim_{n\rightarrow\infty}n\mu_n\log \mu_n^{-1}=\lim_{n\rightarrow\infty} \sqrt{n}\varpi_n\left(\frac{1}{2}\log n +\log \varpi_n^{-1}\right)=\infty.\label{eq:constraint_nlogn}
\end{align}
Assume that $\varpi_n\in O(\frac{1}{\sqrt{n}\log n})$. Then, there exists $0<A<\infty$ such that, for all $n$ large enough, $\varpi_n\leq\frac{A}{\sqrt{n}\log n}$. Since $x\mapsto x\log\frac{1}{x}$ is increasing for $x\in[0,1/e]$, we must have for all $n$ large enough
\begin{align*}
  \lim_{n\rightarrow\infty} \sqrt{n}\varpi_n\left(\frac{1}{2}\log n +\log \varpi_n^{-1}\right) \leq \frac{A}{2} + \lim_{n\rightarrow\infty}\frac{A}{\log n}\left(\frac{1}{2}\log n +\log \log n\right) = A<\infty.
\end{align*}
This contradicts~(\ref{eq:constraint_nlogn}), therefore $\varpi_n\in \omega(\frac{1}{\sqrt{n}\log n})$. 
Consequently, 
\begin{align}
         \lim_{n\rightarrow\infty}\frac{\log M}{\sqrt{n\avgD{{\hQ}^n}{Q_0^\pn}}\log n}&\leq \lim_{n\rightarrow\infty} \frac{n   \avgI{\smash{\tilde{X};\tilde{Y}}}+\Hb{\epsilon_n}}{\sqrt{n^2\avgD{Q_{\mu_n}}{Q_0}}\log n}\\
                                                                                      &\leq \lim_{n\rightarrow\infty} \frac{ \mu_n \kappa \log \mu_n^{-1}-\log(1-\mu_n) +\mu_n \sum_{y\in\calY\setminus\calS} P_{1}(y)\log\frac{P_1(y)}{P_0(y)}+\frac{1}{n}\Hb{\epsilon_n}}{\sqrt{\frac{1}{2}\mu_n^2 \chid{Q_1}{Q_0}(1-\sqrt{\mu_n})}\log n}\\
                                                                                      &=\lim_{n\rightarrow\infty} \frac{  \frac{\kappa}{2} -\kappa \frac{\log \varpi_n}{\log n}-\frac{\log(1-\mu_n)}{\mu_n\log n}+\frac{1}{\log n}\sum_{y\in\calY\setminus\calS} P_{1}(y)\log\frac{P_1(y)}{P_0(y)}+\frac{1}{n\mu_n\log n}\Hb{\epsilon_n}}{\sqrt{\frac{1}{2}\chid{Q_1}{Q_0}(1-\sqrt{\mu_n})}}\\
                                                                                      &= \kappa\sqrt{\frac{2}{\chid{Q_1}{Q_0}}}\left(\frac{1}{2}+\lim_{n\rightarrow\infty}\frac{\log\varpi_n^{-1}}{\log n}\right).
\end{align}
\end{IEEEproof}

Note that Corollary~\ref{cor:scaling_p1_nac_p0} and Theorem~\ref{th:converse_p1_non_ac_p0} differ in the choice of scaling for $\omega_n$ and $\varpi_n$. However, note that for $\varpi_n\in o(1)\cap \omega(\frac{1}{\sqrt{n}\log n})$, we have for all $n$ large enough
\begin{align}
  \frac{\log \varpi_n^{-1}}{\log n}\leq \frac{\log (\sqrt{n}\log n)}{\log n} = \frac{1}{2}+\frac{\log\log n}{\log n},
\end{align}
so that $\lim_{n\rightarrow\infty}\frac{\log\varpi_n^{-1}}{\log n}\leq \frac{1}{2}$. By choosing $\omega_n=n^{\epsilon-1/2}$ for any $\epsilon\in]0;1/2[$ in Corollary~\ref{cor:scaling_p1_nac_p0}, we obtain $\lim_{n\rightarrow\infty}\frac{\log\varpi_n^{-1}}{\log n} =\epsilon$, which can be made arbitrary close to $\frac{1}{2}$. In that regard, the converse is asymptotically tight.

\bibliographystyle{IEEEtran}

\begin{thebibliography}{10}
\providecommand{\url}[1]{#1}
\csname url@samestyle\endcsname
\providecommand{\newblock}{\relax}
\providecommand{\bibinfo}[2]{#2}
\providecommand{\BIBentrySTDinterwordspacing}{\spaceskip=0pt\relax}
\providecommand{\BIBentryALTinterwordstretchfactor}{4}
\providecommand{\BIBentryALTinterwordspacing}{\spaceskip=\fontdimen2\font plus
\BIBentryALTinterwordstretchfactor\fontdimen3\font minus
  \fontdimen4\font\relax}
\providecommand{\BIBforeignlanguage}[2]{{%
\expandafter\ifx\csname l@#1\endcsname\relax
\typeout{** WARNING: IEEEtran.bst: No hyphenation pattern has been}%
\typeout{** loaded for the language `#1'. Using the pattern for}%
\typeout{** the default language instead.}%
\else
\language=\csname l@#1\endcsname
\fi
#2}}
\providecommand{\BIBdecl}{\relax}
\BIBdecl

\bibitem{Bloch2015}
M.~R. Bloch, ``A channel resolvability perspective on stealth communications,''
  in \emph{Proc. of IEEE International Symposium on Information Theory}, Hong
  Kong, June 2015, pp. 2535--2539.

\bibitem{Bash2013}
B.~Bash, D.~Goeckel, and D.~Towsley, ``Limits of reliable communication with
  low probability of detection on {AWGN} channels,'' \emph{{IEEE} {J}ournal on
  {S}elected {A}reas in {C}ommunications}, vol.~31, no.~9, pp. 1921--1930,
  September 2013.

\bibitem{Bash2015a}
B.~A. Bash, A.~H. Gheorghe, M.~Patel, J.~L. Habif, D.~Goeckel, D.~Towsley, and
  S.~Guha, ``Quantum-secure covert communication on bosonic channels,''
  \emph{Nature Communications}, vol.~6, pp.~--, October 2015.

\bibitem{Wang2015}
L.~Wang, G.~Wornell, and L.~Zheng, ``Limits of low-probability-of-detection
  communication over a discrete memoryless channel,'' in \emph{Proc. of IEEE
  International Symposium on Information Theory}, Hong Kong, June 2015, pp.
  2525--2529.

\bibitem{Wang2015a}
\BIBentryALTinterwordspacing
L.~Wang, G.~W. Wornell, and L.~Zheng, ``Fundamental limits of communication
  with low probability of detection,'' arXiv preprint, June 2015. [Online].
  Available: \url{http://arxiv.org/pdf/1506.03236v1.pdf}
\BIBentrySTDinterwordspacing

\bibitem{Ker2007}
A.~D. Ker, ``A capacity result for batch steganography,'' \emph{{IEEE} {S}ignal
  {P}rocessing {L}etters}, vol.~14, no.~8, pp. 525--528, 2007.

\bibitem{Che2013}
P.~H. Che, M.~Bakshi, and S.~Jaggi, ``Reliable deniable communication: Hiding
  messages in noise,'' in \emph{Proc. of IEEE International Symposium on
  Information Theory}, Istanbul, Turkey, July 2013, pp. 2945--2949.

\bibitem{Che2014}
P.~H. Che, M.~Bakshi, C.~Chan, and S.~Jaggi, ``Reliable, deniable and hidable
  communication,'' in \emph{Proc. of Information Theory and Applications
  Workshop}, San Diego, CA, February 2014.

\bibitem{Che2014a}
------, ``Reliable deniable communication with channel uncertainty,'' in
  \emph{Proc. of IEEE Information Theory Workshop}, Hobart, Tasmania, November
  2014, pp. 30--34.

\bibitem{Lee2014}
S.~Lee, R.~Baxley, J.~McMahon, and R.~Frazier, ``Achieving positive rate with
  undetectable communication over {MIMO} {R}ayleigh channels,'' in \emph{Proc.
  of {IEEE} 8th Sensor Array and Multichannel Signal Processing Workshop}, A
  Cor\~una, Spain, June 2014, pp. 257--260.

\bibitem{Lee2014a}
S.~Lee and R.~Baxley, ``Achieving positive rate with undetectable communication
  over {AWGN} and {Rayleigh} channels,'' in \emph{Proc. of IEEE International
  Conference on Communications}, Sydney, Australia, June 2014, pp. 780--785.

\bibitem{Lee2015}
S.~Lee, R.~Baxley, M.~Weitnauer, and B.~Walkenhorst, ``Achieving undetectable
  communication,'' \emph{{S}elected {T}opics in {S}ignal {P}rocessing, {IEEE}
  {J}ournal of}, vol.~9, no.~7, pp. 1195--1205, Oct 2015.

\bibitem{bash2014}
B.~Bash, D.~Goeckel, and D.~Towsley, ``{LPD} communication when the warden does
  not know when,'' in \emph{Proc. IEEE International Symposium on Information
  Theory}, Honolulu, Hawaii, July 2014, pp. 606--610.

\bibitem{Goeckel2016}
D.~Goeckel, B.~Bash, S.~Guha, and D.~Towsley, ``Covert communications when the
  warden does not know the background noise power,'' \emph{IEEE Communications
  Letters}, vol.~PP, no.~99, p.~1, 2016.

\bibitem{Hou2014}
J.~Hou and G.~Kramer, ``Effective secrecy: Reliability, confusion and
  stealth,'' in \emph{Proc. of IEEE International Symposium on Information
  Theory}, Honolulu, HI, July 2014, pp. 601--605.

\bibitem{Bloch2011e}
M.~R. Bloch and J.~N. Laneman, ``Strong secrecy from channel resolvability,''
  \emph{{IEEE} {T}ransactions on {I}nformation {T}heory}, vol.~59, no.~12, pp.
  8077--8098, December 2013.

\bibitem{Che2014c}
P.~H. Che, S.~Kadhe, M.~Bakshi, C.~Chan, S.~Jaggi, and A.~Sprintson,
  ``Reliable, deniable and hidable communication: A quick survey,'' in
  \emph{Proc. of IEEE Information Theory Workshop}, Hobart, Tasmania, November
  2014, pp. 227--231.

\bibitem{Bash2015}
B.~A. Bash, D.~Goeckel, D.~Towsley, and S.~Guha, ``Hiding information in noise:
  fundamental limits of covert wireless communication,'' \emph{IEEE
  Communications Magazine}, vol.~53, no.~12, pp. 26--31, Dec. 2015.

\bibitem{Hero2003}
A.~O. Hero, ``Secure space-time communication,'' \emph{{IEEE} {T}ransactions on
  {I}nformation {T}heory}, vol.~49, no.~12, pp. 3235--3249, December 2003.

\bibitem{Han1993}
T.~S. Han and S.~Verd\'u, ``Approximation theory of output statistics,''
  \emph{{IEEE} {T}ransactions on {I}nformation {T}heory}, vol.~39, no.~3, pp.
  752--772, May 1993.

\bibitem{InformationSpectrumMethods}
T.~S. Han, \emph{{I}nformation-{S}pectrum {M}ethods in {I}nformation
  {T}heory}.\hskip 1em plus 0.5em minus 0.4em\relax Springer, 2002.

\bibitem{Lehmann2005}
E.~Lehmann and J.~Romano, \emph{Testing Statistical Hypotheses}.\hskip 1em plus
  0.5em minus 0.4em\relax Springer, 2005.

\bibitem{ConcentrationInequalities}
S.~Boucheron, G.~Lugosi, and P.~Massart, \emph{Concentration
  inequalities}.\hskip 1em plus 0.5em minus 0.4em\relax Oxford University
  Press, 2013.

\bibitem{TopicsMultiUserIT}
G.~Kramer, \emph{{T}opics in {M}ulti-{U}ser {I}nformation {T}heory}, ser.
  Foundations and Trends in Communications and Information Theory.\hskip 1em
  plus 0.5em minus 0.4em\relax NOW Publishers, 2008, vol.~4, no. 4-5.

\bibitem{Verdu1994}
S.~Verd\'u and T.~S. Han, ``A general formula for channel capacity,''
  \emph{{IEEE} {T}ransactions on {I}nformation {T}heory}, vol.~40, no.~4, pp.
  1147--1157, July 1994.

\bibitem{Cuff2013}
P.~Cuff, ``Distributed channel synthesis,'' \emph{{IEEE} {T}ransactions on
  {I}nformation {T}heory}, vol.~59, no.~11, pp. 7071--7096, 2013.

\bibitem{Hou2014a}
J.~Hou, ``Coding for relay networks and effective secrecy for wire-tap
  channels,'' Ph.D. dissertation, Technischen Universit\"at M\"unchen, 2014.

\bibitem{Bellare2012}
M.~Bellare, S.~Tessaro, and A.~Vardy, ``Semantic security for the wiretap
  channel,'' in \emph{Advances in Cryptology - CRYPTO 2012}, ser. Lecture Notes
  in Computer Science, R.~Safavi-Naini and R.~Canetti, Eds., vol. 7417.\hskip
  1em plus 0.5em minus 0.4em\relax Springer Berlin Heidelberg, 2012, pp.
  294--311.

\bibitem{Han2014}
T.~S. Han, H.~Endo, and M.~Sasaki, ``Reliability and secrecy functions of the
  wiretap channel under cost constraint,'' \emph{{IEEE} {T}ransactions on
  {I}nformation {T}heory}, vol.~60, no.~11, pp. 6819--6843, 2014.

\bibitem{Renes2011}
J.~Renes and R.~Renner, ``Noisy channel coding via privacy amplification and
  information reconciliation,'' \emph{{IEEE} {T}ransactions on {I}nformation
  {T}heory}, vol.~57, no.~11, pp. 7377--7385, 2011.

\bibitem{Cachin2004}
C.~Cachin, ``An information-theoretic model for steganography,''
  \emph{{I}nformation and {C}omputation}, vol. 192, no.~1, pp. 41--56, July
  2004.

\bibitem{Lin1991a}
J.~Lin, ``Divergence measures based on the shannon entropy,'' \emph{{IEEE}
  {T}ransactions on {I}nformation {T}heory}, vol.~37, no.~1, pp. 145--151,
  January 1991.

\bibitem{Polyanskiy2010}
Y.~Polyanskiy, H.~V. Poor, and S.~Verd\'u, ``Channel coding rate in the finite
  blocklength regime,'' \emph{{IEEE} {T}ransactions on {I}nformation {T}heory},
  vol.~56, no.~5, pp. 2307--2359, May 2010.

\end{thebibliography}

\begin{IEEEbiography}[{\includegraphics[width=1in,height=1.25in,clip,keepaspectratio]{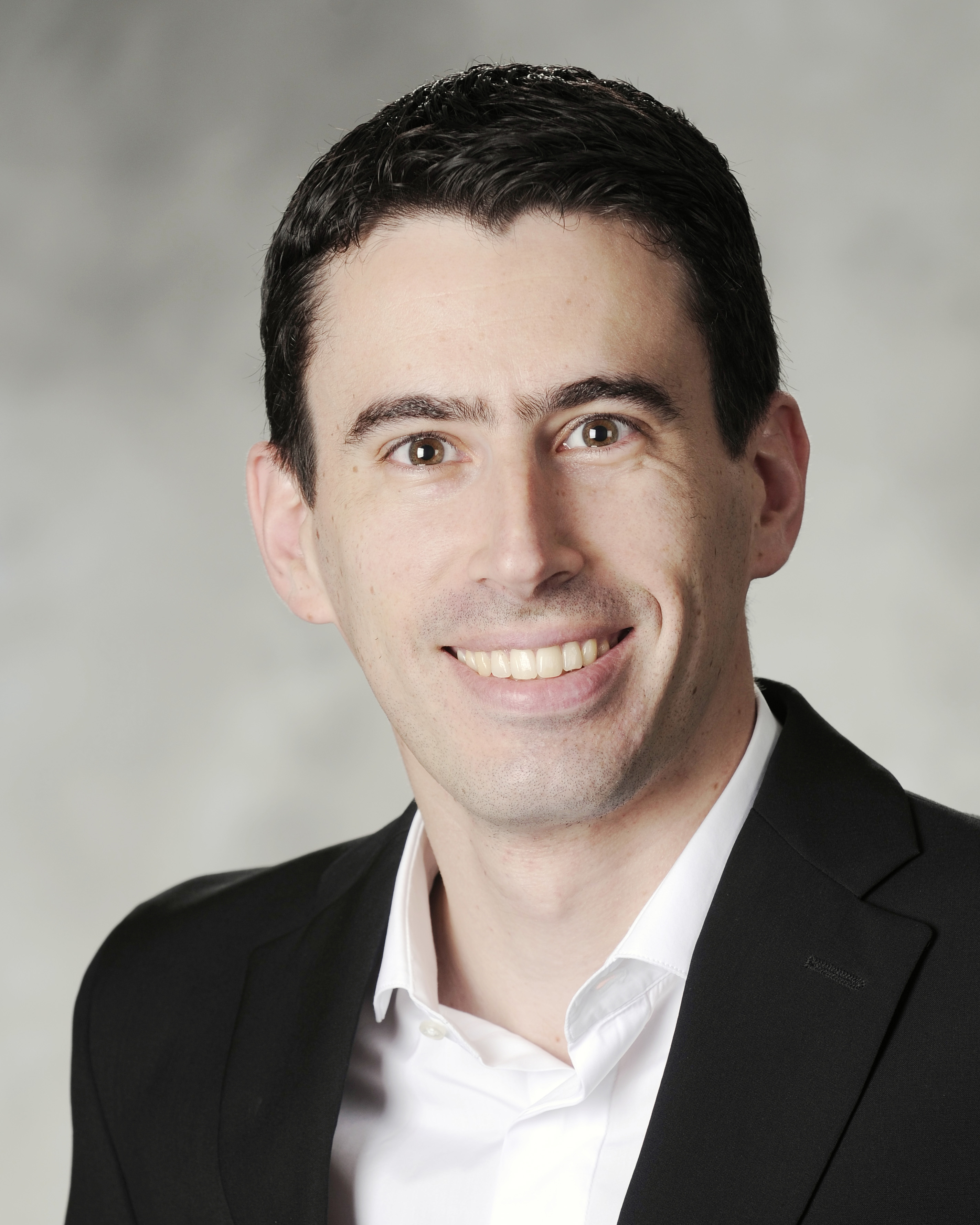}}]{Matthieu Bloch}
is an Associate Professor in the School of Electrical and Computer Engineering. He received the Engineering degree from Sup\'elec, Gif-sur-Yvette, France, the M.S. degree in Electrical Engineering from the Georgia Institute of Technology, Atlanta, in 2003, the Ph.D. degree in Engineering Science from the Universit\'e de Franche-Comt\'e, Besan\c{c}on, France, in 2006, and the Ph.D. degree in Electrical Engineering from the Georgia Institute of Technology in 2008. In 2008-2009, he was a postdoctoral research associate at the University of Notre Dame, South Bend, IN. Since July 2009, Dr. Bloch has been on the faculty of the School of Electrical and Computer Engineering, and from 2009 to 2013 Dr. Bloch was based at Georgia Tech Lorraine. His research interests are in the areas of information theory, error-control coding, wireless communications, and cryptography. Dr. Bloch is a member of the IEEE and has served on the organizing committee of several international conferences; he was the chair of the Online Committee of the IEEE Information Theory Society from 2011 to 2014, and he has been on the Board of Governors of the IEEE Information Theory Society since January 2016. He is the co-recipient of the IEEE Communications Society and IEEE Information Theory Society 2011 Joint Paper Award and the co-author of the textbook \emph{Physical-Layer Security: From Information Theory to Security Engineering} published by Cambridge University Press.
\end{IEEEbiography}
\end{document}